\documentclass[man, floatsintext]{apa7}

\usepackage{amsmath, amssymb, mathptmx, enumitem, array, enumitem, bm}
\usepackage{tabularx, makecell, setspace, courier, bm}
\usepackage{algorithm, algpseudocode}
\usepackage{caption}
\usepackage[american]{babel}
\usepackage{etoolbox}
\usepackage[figuresleft]{rotating}
\usepackage[style=apa,sortcites=true,sorting=nyt,backend=biber]{biblatex}

\DeclareLanguageMapping{american}{american-apa} 
\DeclareMathAlphabet{\cmcal}{OMS}{cmsy}{m}{n}
\graphicspath{ {./figures} } 
\DeclareMathOperator*{\vect}{vec}
\algrenewcommand\algorithmicensure{\textbf{Return:}}

\newlength{\algcolwidth}
\newcommand{\alignedfor}[2]{%
  \makebox[\algcolwidth][l]{#1}%
  #2%
}
\newcommand{\N}[3][]{N_{#1} \left(#2,\: #3\right)}
\newcommand{\GP}[3][]{GP_{#1} \left(#2,\: #3\right)}
\newcommand{\IW}[3][]{IW_{#1} \left(#2,\: #3\right)}
\newcommand{\IG}[2]{IG \left(#1,\: #2\right)}
\newcommand{\MN}[4][]{MN_{#1} \left(#2,\: #3,\: #4\right)}
\newcommand{\p}[1]{p \left(#1 \right)}
\newcommand{\q}[2][]{q_{#1} \left(#2 \right)}
\newcommand{\Pbb}[1]{\mathbb{P} \left(#1 \right)}
\newcommand{\pcond}[2]{p\left(#1\: \mid\: #2\right)}
\newcommand{\qcond}[3][t]{q_{#1} \left(#2\: \mid\: #3\right)}
\newcommand{\vcond}[3][t-1]{v_{#1} \left(#2\: \mid\: #3\right)}

\newcommand{\fdet}[2][]{\left| #2 \right|^{#1}}
\newcommand{\fgamma}[2][]{\Gamma_{#2} \left(#1\right)}

\newcommand{\x}[1][t]{\mathbf{x}_{#1}}
\newcommand{\y}[1][t]{\mathbf{y}_{#1}}
\newcommand{\YT}[1][1:T]{\mathbf{Y}_{#1}}
\newcommand{\XT}[1][1:T]{\mathbf{X}_{#1}}

\newcommand{\epsilont}[1][t]{\mathbf{\epsilon}_{#1}}
\newcommand{\etat}[1][t]{\mathbf{\eta}_{#1}}
\newcommand{\Sigmaeta}[1][\eta]{\bm{\Sigma}_{#1}}
\newcommand{\Sigmaepsilon}[1][\epsilon]{\bm{\Sigma}_{#1}}
\newcommand{\Sigmapostcol}{\bm{\Sigma}'_{\bm{S}_{\hpars}}}

\newcommand{\vprioreta}[1][\eta]{v_{#1}}
\newcommand{\vposteta}[1][\eta]{v'_{#1}}
\newcommand{\Psiprioreta}[1][\eta]{\bm{\Psi}_{#1}}
\newcommand{\Psiposteta}[1][\eta]{\bm{\Psi}'_{#1}}

\newcommand{\mupriorlambda}[1][\lambda]{\bm{\mu}_{#1}}
\newcommand{\mupostlambda}[1][\lambda]{\bm{\mu}'_{#1}}
\newcommand{\Sigmapriorlambda}[1][\lambda]{\bm{\Sigma}_{#1}}
\newcommand{\Sigmapostlambda}[1][\lambda]{\bm{\Sigma}'_{#1}}
\newcommand{\vpriorepsilon}[1][\epsilon]{v_{#1}}
\newcommand{\vpostepsilon}[1][\epsilon]{v'_{#1}}
\newcommand{\Psipriorepsilon}[1][\epsilon]{\bm{\Psi}_{#1}}
\newcommand{\Psipostepsilon}[1][\epsilon]{\bm{\Psi}'_{#1}}

\newcommand{\identity}[1][]{\bm{I}_{#1}}
\newcommand{\hpars}{\bm{\theta}}
\newcommand{\ds}{d}
\newcommand{\dobs}{o}

\newcommand{\loadmat}{\bm{\Lambda}}
\newcommand{\muinit}{\bm{\mu}_{1}}
\newcommand{\Sigmainit}{\bm{\Sigma}_{1}}
\newcommand{\Sigmaout}{\bm{\Sigma_\ds}}

\newcommand{\hidx}{h}

\newcommand{\skernel}[3][]{k_{#1}\!\left(\: \left| #2 - #3 \right| \:\right)}

\newcolumntype{s}{>{\hsize=.6\hsize}X}

\AtBeginEnvironment{tabularx}{\setlist[itemize, 1]{wide,
    leftmargin=*, itemsep=0pt, before=\vspace{-\dimexpr\baselineskip +2
      \partopsep}, after=\vspace{-\baselineskip}}}

\title{Learning Nonlinear Dynamics: Improving the Estimation Efficiency and Reliability of Gaussian Process State-Space Models}

\shorttitle{Improving the Estimation of Gaussian Process State-Space Models}

\leftheader{Failenschmid}

\authorsnames{{Jan I. Failenschmid}, {Leonie V.D.E. Vogelsmeier}, {Joris
      Mulder}, {Joran Jongerling}}

\authorsaffiliations{{Tilburg University}}

\abstract{Understanding dynamic systems is a central goal in many scientific
  disciplines. State-space models provide a general framework for studying
  latent dynamic systems based on indirect observations. 
  However, classical state-space methods require researchers to specify the
  parametric form of the system dynamics in advance, which can be challenging
  when the underlying processes are nonlinear and only partially
  explained by theory. 
  Gaussian process state-space models address this by
  learning the system dynamics directly from data. However, estimating these models exactly can become computationally infeasible for
  moderately long time-series. In this paper, we therefore aim to improve the Bayesian estimation
  of approximate Gaussian process state-space models to make these models more accessible and facilitate the
  statistical learning of nonlinear dynamic systems in empirical research.
  To this end, we first propose two modifications to an existing Gibbs sampler
  for these models
  that considerably improve its sampling efficiency and convergence. Second, we
  use a confirmatory factor analysis measurement model, which reduces   identifiability issues and allows researchers to impose a specific measurement
  structure on the model. Third, we provide a systematically validated software
  implementation of the model and sampler for applied use in empirical research.
  To validate the sampler, we conducted a simulation-based calibration which
  showed that the sampler converged reliably across many simulated
  data sets and produces well-calibrated posterior inferences. We further
  illustrate how the model can be applied and interpreted using two empirical
  examples. Together, these contributions provide a practical and validated
  workflow for learning nonlinear latent dynamics with Gaussian process
  state-space models. }

\keywords{state space model, Gaussian process, Hilbert space methods, confirmatory factor analysis, Gibbs sampling, particle filtering
}

\authornote{ 
  \addORCIDlink{Jan I. Failenschmid}{0009-0007-5106-7263} \\
  \addORCIDlink{Leonie V.D.E. Vogelsmeier}{0000-0002-1666-7112} 

  Correspondence concerning this article should be addressed to Jan I.
  Failenschmid, Tilburg School of Social and Behavioral Sciences: Department o
  Methodology and Statistics, Tilburg University, Warandelaan 2, 5037A  Tilburg,
  Netherlands. E-mail: J.I.Failenschmid@tilburguniversity.edu 

  \noindent \textbf{Data and code availability statement:} \\
  \noindent  All code used in this paper is available in the online
  supplementary material and can be used to reproduce the simulated data. The
  data used in the empirical examples are available through the \texttt{astsa} R
  package \parencite{shumwayTimeSeriesAnalysis2025}.

  \noindent \textbf{Disclosure of artificial intelligence-generated content:} \\
  \noindent Generative artificial intelligence tools were used to improve the
  spelling, grammar, and overall editing of this manuscript. They were also used
  for code review, code editing, and automated testing. No text or code was
  independently generated by artificial intelligence tools, and all presented
  writing and code reflect the authors' original work.

  \noindent \textbf{Acknowledgements:}\\
  \noindent This project was supported by a Starter Grant. Additionally, J.M.
  was supported by an ERC Consolidator Grant (101087383).

  \newpage

  \noindent \textbf{Author contribution:}\\
  \noindent \textbf{Jan I. Failenschmid}: Writing - Original Draft Preparation,
  Software, Formal Analysis, Data Curation, Conceptualization, Investigation,
  Methodology, Visualization. \textbf{Leonie V.D.E. Vogelsmeier}: Writing -
  Original Draft Preparation, Writing - Review \& Editing, Supervision,
  Conceptualization, Funding Acquisition, Project Administration,
  Investigation, Methodology. \textbf{Joris Mulder}: 
  Writing - Original Draft Preparation, Writing - Review \&
  Editing, Supervision, Conceptualization, Project Administration, 
  Investigation, Methodology.
  \textbf{Joran Jongerling}: Writing - Original Draft Preparation, Writing -
  Review \& Editing, Supervision, Conceptualization, Funding Acquisition,
  Project Administration, Investigation, Methodology.
  }

\begin{document}

\maketitle

Understanding how systems evolve over time is a central goal across many
scientific disciplines. Ecologists, for example, study the dynamics of
interacting predator and prey populations
\parencite{murrayMathematicalBiology1993}, psychologists examine fluctuations in
well-being and affect across days or weeks \parencite{waughAffectDynamics2021},
economists investigate changes in financial markets and business cycles
\parencite{zengStateSpaceModelsApplications2013}, and neuroscientists seek to
characterize the temporal evolution of neural activity
\parencite{fristonNonlinearResponsesFMRI2000}. Across these domains, researchers
are not only interested in how variables evolve over time, but also in how their
mutual influences on one another give rise to complex patterns of change that
cannot be understood by studying each variable in isolation
\parencite{vandermaasComplexSystemsResearchPsychology2024}.
A useful way to study such questions is to view the relevant variables and the
processes governing their change as a dynamic system
\parencite{bokerDynamicalSystemsDifferential2012}. Here, we refer to these
processes as the system dynamics. Formally, a dynamic system consists of one or
more variables that change interdependently over time, such that their
values at a given time point can be modelled depending on their
previous values. One of the best-known examples of such a dynamic system, which
we will use as a running example throughout this paper, comes from ecology and
concerns the coupled population dynamics of a predator and a prey species
\parencite{murrayMathematicalBiology1993}.
In this example, the dynamics are commonly described by the
\textit{Lotka-Volterra} equations, which
state that the prey population is intrinsically growing but starts to decline as
predation increases and the predator population naturally declines but starts to
grow when prey are sufficiently abundant
\parencite{murrayMathematicalBiology1993,
wangerskyLotkaVolterraPopulationModels1978}.

State-space models (SSMs) provide a statistical framework for modelling such
dynamic systems and analyzing time-series data
\parencite{durbinTimeSeriesAnalysis2012}, with applications ranging from
engineering \parencite{friedlandControlSystemDesign2005}, physics
\parencite{abarbanelStatisticalPhysicsData2022,evensenDataAssimilationFundamentals2022},
and ecology \parencite{newmanStatespaceModelsEcological2023} to economics
\parencite{zengStateSpaceModelsApplications2013} and psychology
\parencite{chowEquivalenceDifferencesStructural2010,driverContinuousTimeStructural2017}.
Within an SSM, the variables that constitute the system (e.g., the predator and
prey population sizes) are referred to as the state variables, while the system
dynamics are represented through a dynamic (or transition) model, which
describes how the current values of the state variables depend on the previous
ones. Since, in this paper, we focus on discrete-time SSMs an appropriate
dynamic model for the predator-prey example could correspond to a discrete-time
version of the \textit{Lotka-Volterra} equations
\parencite{murrayMathematicalBiology1993,linDynamicsChaosControl2020,eskandariDynamicsBifurcationsDiscretetime2025}.

In addition to this, SSMs typically treat the state variables as latent
variables that cannot be observed directly
\parencite{durbinTimeSeriesAnalysis2012}. Instead, they are measured indirectly
through a set of observed indicator variables. These could, for instance, be
capture-recapture counts in ecology or, in the classical predator-prey example,
even more indirectly the number of pelts sold from each species. Therefore, SSMs
include a measurement model that describes how the observations depend on the
state variables. In the predator-prey example, the measurement model would
describe how the number of sold pelts for the predator and prey species depends
on their underlying population sizes \parencite{durbinTimeSeriesAnalysis2012}.

Statistical methods for SSMs classically focus on estimating the parameters of
the dynamic and measurement models
\parencite{durbinTimeSeriesAnalysis2012, sarkkaBayesianFilteringSmoothing2023}.
A wide range of methods for doing so has been developed for both linear and
nonlinear SSMs. In a linear SSM, the dynamic and measurement models are linear
functions of the latent states, whereas in nonlinear SSMs one or both of
these functions are nonlinear. Implementations of these methods are available
in several statistical software packages. For example, in R \parencite{R-base},
such methods are implemented in packages such as \texttt{dynr}
\parencite{ouWhatsDynrPackage2019} and \texttt{ctsem}
\parencite{driverContinuousTimeStructural2017}. However, these methods require
researchers to specify the functional forms of the dynamic and measurement
models in advance. These could, for instance, be informed by substantive
theories about the underlying dynamic system and the measurement process. As
described before, in the predator-prey example the dynamic model could
correspond to a discrete time version of the \textit{Lotka-Volterra}
equations. However, when such substantive theories are missing or lack
precision, specifying the appropriate type of change can be challenging. In
this case, researchers often default to using a linear dynamic model. However,
in many empirical applications, the underlying dynamic system is unlikely to be
exactly linear. Consequently, the resulting linear SSM may not be able to
accurately capture important behaviors of the system or patterns in the data
that emerge from the system's nonlinear dynamics.  
For instance, a linear SSM would not be able to capture the critical interaction
effects between the two species in the predator-prey example.

To address this, several methods have been developed that learn some or all
components of an SSM directly from the data. This allows researchers to impose
fewer structural and parametric assumptions on the dynamic system and
measurement process. One such approach is the Gaussian process state-space model
\parencite[GP-SSM;
][]{frigolaBayesianInferenceLearning2013,turnerStateSpaceInferenceLearning2010},
which uses Gaussian processes (GP), a Bayesian machine learning technique
\parencite{rasmussenGaussianProcessesMachine2006}, to learn the form of the dynamic model, the measurement model, or both simultaneously. 

There are several reasons why GPs are a particularly well suited machine
learning method for being used within SSMs, First, a GP constitutes a Bayesian
prior over a space of candidate functions and learning a function with a GP is
equivalent to obtaining its posterior distribution
\parencite{rasmussenGaussianProcessesMachine2006}. This provides a principled
way to account for the additional epistemic uncertainty that is created by
learning the form of some components of the SSM from data, rather than assuming
that their functional form is known a priori and that only a fixed set of
parameters is estimated. Second, GP priors are very flexible and can be used to
encode assumptions and prior expectations about certain properties of the target
function, such as its smoothness, average behavior, or periodicity.
Additionally, multiple GP priors can be combined to create a more flexible model
and learn more complex functions
\parencite{duvenaudStructureDiscoveryNonparametric}. Lastly, GPs are capable of
modelling the functional relationships between multiple latent variables
\parencite{mukherjeeHilbertSpaceMethods2025}, such as the state variables in an
SSM.

The primary drawback of using GPs, however, is their computational cost, which
scales cubically with the number of observed data points (i.e., exact GP
inference requires inverting a covariance matrix over all observed data points).
This quickly makes inferring the exact GP computationally demanding, even for
moderately long time-series. To address this, several approximation methods have
been developed for GPs. One of them is the Hilbert space or reduced-rank GP
approximation, which represents the GP using a finite set of basis functions
\parencite{solinHilbertSpaceMethods2019}.
\textcite{svenssonComputationallyEfficientBayesian2016} show that this basis
function approximation can be estimated efficiently within a GP-SSM, since it
corresponds to a standard Bayesian regression problem, which is computationally
more efficient when the number of basis functions is low and converges to the
exact GP as the number of basis functions goes to infinity, under certain
regularity conditions. They further show, that the Hilbert space approximate
GP-SSM (HSGP-SSM), outperforms alternative sparse approximations and the exact
GP in terms of both computation time and prediction accuracy on a simulated
benchmark dataset\footnote{Additionally,
\textcite{mukherjeeHilbertSpaceMethods2025} showed that HSGPs might be estimated
more accurately than the exact GP when using latent variables as input points.
This could make them particularly suitable for being used within SSMs.}. Because
of this, HSGP-SSMs are a promising method for studying and learning dynamic
systems. 

However, there are currently several challenges regarding the Bayesian
estimation of HSGP-SSMs which limit their applicability and hinder their
wide-spread use for learning dynamic processes in empirical research. Therefore,
this paper focuses on improving the estimation of HSGP-SSMs via MCMC sampling.
By doing so, we aim to make this method more accessible and widely applicable
for applied science and facilitate the statistical learning of nonlinear dynamic
systems using a flexible data-driven approach building on non-parametric GPs.
Specifically, we identified three challenges, which currently make using
HSGP-SSMs more complicated in practice. Below, we describe each of these
challenges and how we aim to address them.

First, existing MCMC algorithms for HSGP-SSMs can have a low sampling efficiency
for certain model parameters. The Gibbs sampler proposed by
\textcite{svenssonComputationallyEfficientBayesian2016} 
can accurately estimate both the HSGP and the latent states of the SSM. However,
Gaussian processes also include hyperparameters that control key properties of
the GP prior, such as the overall variability of the functions it describes and
how rapidly they can change across the input space. In doing so, they directly
influence the types of functions that the GP can learn. In a fully Bayesian
framework, these hyperparameters are estimated jointly with the remaining model.
For the hyperparameters, the Gibbs sampler has a low sampling efficiency,
which often leads to slow mixing and convergence of the MCMC chains
(Figure~\ref{fig:orig_trace}). This also negatively affects the estimation of
the remaining model parameters. To address this issue, we propose two
modifications to their algorithm aimed at improving its sampling efficiency and
facilitating more reliable estimation in empirical applications.

\begin{figure}[!ht]
  \caption{Comparison of hyperparameter trace plots from
  \textcite{svenssonComputationallyEfficientBayesian2016} and the proposed
  sampler}
  \fitfigure{combined_trace.png}
  \figurenote{This figure compares the sampler of
  \textcite{svenssonComputationallyEfficientBayesian2016} with the sampler
  proposed in this paper. The upper panels show the original hyperparameter
  trace plots for the first synthetic data example of
  \textcite{svenssonComputationallyEfficientBayesian2016} after removing the
  burn-in. In particular, the marginal variance parameter shows very
  slow mixing. The bulk effective sample size is $6.85$ for the marginal
  variance and $56.5$ for the lengthscale. The tail effective
  sample sizes are $22.2$ and $98.6$, respectively.    
  The lower panels show trace plots produced by the sampler proposed in this
  paper for the same dataset, using the same priors and settings. However, in addition to the GP, we also estimated the dynamic and measurement
  error covariances. The bulk effective sample size increases to $291$ for the
  marginal variance and $106$ for the lengthscale. The corresponding tail
  effective sample sizes are $501$ and $124$.}
  \label{fig:orig_trace}
\end{figure}

The second challenge in GP-SSMs is identifiability. When both the dynamic and
measurement models are estimated with too much flexibility, the model becomes
weakly identified, as multiple combinations of functions can explain the
observed data equally well. During MCMC sampling this can lead to drifting,
numerical instability, and difficulties in assessing the convergence of the sampler. Additionally,
it can complicate the interpretation of the resulting posterior distributions
\parencite{gelfandIdentifiabilityImproperPriors1999}. To mitigate these issues,
it is common in the GP-SSM literature to focus on learning the dynamic model
while assuming a known or parametric measurement model
\parencite{svenssonComputationallyEfficientBayesian2016,
eleftheriadisIdentificationGaussianProcess,
frigolaBayesianInferenceLearning2013}\footnote{A notable exception is
\textcite{yuGaussianProcessFactorAnalysis2009}, who learn the measurement model
non-parametrically while assuming a linear dynamic model.}. We therefore use a
confirmatory factor analysis (CFA) model as the measurement
model. The CFA is a widely used
linear\footnote{Assuming a linear measurement model does not reduce the
generality of the model, because any GP-SSM of the form given in
Equation~\ref{eq:ssm_base} can be reparametrized as a model with a linear
measurement function through state expansion
\parencite{frigola-alcaldeBayesianTimeSeries2015}.} measurement model from structural equation modeling
\parencite{muthenBayesianStructuralEquation2012,
brownConfirmatoryFactorAnalysis2015}, which can be identified through parameter
constraints \parencite{hayashiExaminingIdentificationIssues2006}. These
constraints also allow researchers to impose a specific structure on the
measurement model, which is important as each state is often measured only by a
subset of the observed variables in empirical applications
\parencite{brownConfirmatoryFactorAnalysis2015,
floydFactorAnalysisDevelopment1995}.

Lastly, there is currently no validated software implementation for estimating
HSGP-SSMs. Researchers thus need to implement their own estimators or
adapt published code, which has often only been evaluated on a limited set of
simulated datasets. This is both inefficient and prone to human error. It is
therefore important to have an accessible and systematically validated
implementation of an HSGP-SSM estimator. To this end, we evaluate the sampler
proposed in this paper over a wide range of simulated data sets using
simulation-based calibration
\parencite[SBC;][]{modrakSimulationBasedCalibrationChecking2022}. This is a
simulation-based method for empirically testing whether a Bayesian sampler
produces valid samples from the target posterior distribution.

In the following, we introduce the HSGP-SSM in detail (Section~\ref{methods}).
We then present the Gibbs sampler for this model, along with our proposed
modifications to improve its efficiency (Section~\ref{estimation}), and validate
the sampler using simulation-based calibration (Section~\ref{sbc}). Lastly, we
illustrate the HSGP-SSM on two empirical data sets (Section~\ref{emp_example}).

\setcounter{section}{1}
\section{Gaussian process state space model}\label{methods}

\subsection{The State-Space Model}

We consider the following SSM \parencite{durbinTimeSeriesAnalysis2012}, which describes a system of $\ds$
unobserved state variables $\x[]$ that are measured through a set of $\dobs$
observed variables $\y[]$ at the time points $t = {1, \dots, T}$.

\begin{equation}\label{eq:ssm_base}
  \begin{aligned}
    \x[t+1] & = f(\x) + \etat, \qquad           & \etat \sim \N[]{\bm{0}}{
    \Sigmaeta}                                                               \\
    \y      & = \loadmat \x + \epsilont, \qquad & \epsilont \sim \N{\bm{0}}{
      \Sigmaepsilon}
  \end{aligned}
\end{equation}

\noindent This model contains the dynamic transition function $f$, which models
the system dynamics by describing how the latent states evolve from one time
point to the next. In a linear model, $f$ would contain the autoregressive and
cross-lagged effects of the state variables on one another. More generally, if
$f$ is nonlinear, it captures all nonlinear effects and dependencies between the
state variables. In the predator-prey example, $f$ is a multivariate nonlinear
function that describes the tendency of the prey population to grow in the
absence of predation, the tendency of the predator population to decline in the
absence of prey, and all ways in which the two populations influence each other
over time \parencite{murrayMathematicalBiology1993}. However, specifying an
appropriate parametric form for $f$ as a function of the predator and prey
population sizes can be challenging. Moreover, the true
data-generating process will rarely follow the assumed parametric form exactly.
When the chosen model closely approximates reality, the resulting inference may
still be adequate. However, substantial misspecification can lead to poor
inference and erroneous conclusions. A key contribution of GP-SSMs is the
ability to learn the form of $f$ directly from the data
\parencite{frigola-alcaldeBayesianTimeSeries2015}. This is particularly useful
in empirical applications, where the system dynamics are likely to be nonlinear
and difficult to specify a priori.  

Since we use a CFA as the measurement model, it is linear and can be represented by the loading matrix $\loadmat$, which contains the factor loadings (i.e., regression coefficients) that relate the latent states to the indicator variables \parencite{muthenBayesianStructuralEquation2012}.
We intentionally omit an
intercept from the measurement model, as it is generally not identified if
the dynamic function $f$ is estimated non-parametrically, because
any constant shift in the latent states could then be absorbed by the
measurement intercept and the dynamic function, producing a model that would
make the data equally likely. The dynamic errors $\etat$ and the measurement
errors $\epsilont$ are assumed to be additive to the system, independent across
time points, and normally distributed with a zero mean. Their covariances are
given by the dynamic error covariance matrix $\Sigmaeta$ and the measurement
error covariance matrix $\Sigmaepsilon$ respectively.

Following common practice \parencite{durbinTimeSeriesAnalysis2012}, we choose a
multivariate normal distribution as a prior distribution for the initial state
of the system

\begin{equation}\label{eq:init_dist}
  \begin{aligned}
    \x[1] & \sim \N{\muinit}{\Sigmainit}.
  \end{aligned}
\end{equation}

\noindent Here, the mean vector $\muinit$ corresponds to our prior expectation
for the initial state $\x[1]$ and the covariance matrix $\Sigmainit$ quantifies
our prior uncertainty about $\x[1]$. These parameters can be chosen based on
substantive knowledge about the state variables. For example, if the state
variables are on a large scale or there is a lot of uncertainty about the
initial state a large prior variance could be used.

\subsection{The Gaussian Process}

In a GP-SSM the possibly nonlinear dynamic function $f$ in
Equation~\ref{eq:ssm_base} is then learned using a GP
\parencite{frigola-alcaldeBayesianTimeSeries2015,frigolaBayesianInferenceLearning2013},
which defines a probability distribution over a space of candidate functions for
$f$. In a Bayesian model, this probability distribution can be used as a prior
distribution for $f$, so that it can be learned by obtaining its respective
posterior distribution conditional on the data
\parencite{rasmussenGaussianProcessesMachine2006}.

Since $f: \mathbb{R}^{\ds} \rightarrow \mathbb{R}^{\ds}$ is a multivariate
function that maps the $\ds$-dimensional state vector to its subsequent value,
we need a multi-output GP with the same input and output dimensions
\parencite{alvarezKernelsVectorValuedFunctions2012}. For instance in the running
predator-prey example, $f$ describes how the current predator and prey
population sizes relate to their values at the next time point. Hence, the GP
would need to define a distribution over functions that take two inputs (e.g.,
the current predator and prey population sizes) and return two outputs (e.g.,
their subsequent population sizes).

We therefore use the multi-output GP prior 

\begin{equation}\label{eq:gp_prior}
  \begin{aligned}
    f &\sim \GP{m_{\hpars}(\x)}{K_{\hpars}(\x, \x')}, \\
    \text{with}\ m_{\hpars}(\x) &= \mathbf{0}, \\
    K_{\hpars}(\x, \x') &= \skernel[\hpars]{\x}{\x'} \Sigmaout .
  \end{aligned}
\end{equation}

\noindent Here, $m_{\hpars}(\x): \mathbb{R}^{\ds} \rightarrow \mathbb{R}^{\ds}$
is a vector-valued mean function, which defines the expected value of each
output dimension of the GP at the input point $\x$. The covariance function
$K_{\hpars}(\x, \x'): \mathbb{R}^{\ds} \times \mathbb{R}^{\ds} \rightarrow
\mathbb{R}^{\ds \times \ds}$ is matrix-valued and specifies the covariance
between the output dimensions of the GP at two input points $\x$ and $\x'$
\parencite{alvarezKernelsVectorValuedFunctions2012}. In order to form a
conjugate prior, we need a covariance function that factorizes into a scalar
kernel $\skernel[\hpars]{\x}{\x'}$, which captures the covariance across input
points, and a output covariance matrix $\Sigmaout$, which captures the
covariance across output dimensions
\parencite{svenssonComputationallyEfficientBayesian2016}\footnote{Gaussian
processes with this factorized covariance structure are also known as intrinsic
coregionalization models.}. The mean and covariance function jointly determine
the shapes of the functions that can be modelled by the GP
\parencite{rasmussenGaussianProcessesMachine2006}. Lastly, the vector $\hpars$
contains the hyperparameters of the GP.

To understand what this prior does, it helps to see the GP as a generalization
of the multivariate normal distribution to continuous input spaces. Whereas a
multivariate normal distribution is defined over a fixed set of dimensions, a GP
is defined over all possible values that its input variables can take, such that
evaluating the GP at any finite set of input points yields a multivariate normal
distribution with one dimension for each input point
\parencite{rasmussenGaussianProcessesMachine2006}.

For notational purposes let $\XT[]$ be a matrix where each column is one vector
valued input point $\x$. We further use $f(\XT[])$ to denote a vector valued
function evaluated on each column of this matrix, such that $f(\XT[]) =
(f(\x[1]), \ldots, f(\x[T]))$. Then, because the covariance function factors
into an output covariance matrix and an input covariance kernel, the prior
distribution for $f$ at a fixed set of input points $\XT[]$ can be written as a
matrix normal distribution\footnote{Note, that this follows directly from the
fact that a multi-output GP can be written as a single output GP 
on an augmented input space $(\x, \ds)$. This can be achieved by stacking the
multiple outputs in a single vector and using an appropriately structured
covariance function $K_{\hpars}((\x, d_i), (\x', d_j)) =
\skernel[\hpars]{\x}{\x'} \sigma_{i,j}$. Evaluated at a
finite set of input points this GP implies $\vect(f(\XT[])) =
\N{\vect(m_{\hpars}(\XT[]))}{k_{\hpars}(\XT[], \XT[]) \otimes \Sigmaout}$.} 

\begin{equation}
  f(\XT[]) = \MN{m_{\hpars}(\XT[])}{\Sigmaout}{
    k_{\hpars}(\XT[], \XT[])},
\end{equation}

\noindent where $k_{\hpars}(\XT[], \XT[])$ denotes a covariance matrix generated
by evaluating the kernel $\skernel[\hpars]{\x}{\x'}$ pairwise on all columns of
$\XT[]$.

\subsubsection{Choosing the mean and covariance function}

In theory any valid mean and covariance function can be used to define a GP
prior\footnote{The covariance function needs to define a symmetric, positive
semi-definite covariance matrix for all sets of input points $\XT[]$ and
$\XT[]'$ over all output dimensions.}. However, we put additional restrictions
on the mean and covariance function, which make it possible to estimate the
model more efficiently \parencite{svenssonComputationallyEfficientBayesian2016}.
Specifically, we set the mean function to zero, which is a common practice and
corresponds to having no prior expectation about systematic deviations from zero
\parencite{rasmussenGaussianProcessesMachine2006}. However, the presented GP-SSM
can also easily be adapted to other mean functions that are either fixed or
constitute a multivariate regression model
\parencite{blightBayesianApproachModel1975}.

The covariance function conceptually determines how much the functions that are
described by the GP vary at each input point, through the variances, and how
(dis)similar they are at different input points $\x$ and $\x'$, through the
covariances, across all output dimensions
\parencite{rasmussenGaussianProcessesMachine2006}. In the predator-prey example,
this would correspond to how similarly the population dynamics behave at
different values for the predator and prey population sizes. If the covariance
function implies a high correlation for two sets of population sizes $\x$ and
$\x'$, then the GP will predict similar values for the population sizes at the
next time point for both input points.

In our covariance function, the output covariance matrix $\Sigmaout$ captures
how similar the functions described by the GP are across the output dimensions.
In the predator-prey example, it would therefore describe how strongly the GP
priors for the predator and prey dynamics are correlated at a given time point.
Since the number of output dimensions is typically small compared to the number
of input points in GP-SSMs, $\Sigmaout$ can be specified without further
restrictions.

The covariance across the input points is modeled using a kernel
$\skernel[\hpars]{\x}{\x'}$ \parencite{rasmussenGaussianProcessesMachine2006},
which defines the covariance between the values that the functions defined by
the GP take at any two input points, irrespective of the output dimension. The
kernel is chosen to be stationary, such that the covariance only depends on the
distance between the input points, which is necessary for the HSGP
approximation. This implies that we have the same prior believes about $f$ at
all input points, which leaves the statistical properties of the GP invariant
across the input space.

The kernel further determines the types of functions that the GP can represent.
It can therefore be chosen to reflect prior assumptions about the shape of $f$,
such as its smoothness or periodicity. For simplicity, we focus on the widely
used squared-exponential kernel, which is appropriate for modeling smooth
nonlinear functions \parencite[i.e., functions with a gradual rate of
change][]{rasmussenGaussianProcessesMachine2006}. The results presented in this
paper can also be applied to other stationary kernels, such as those from the
Matérn family, which are suitable for rougher functions (i.e., functions with an
abrupt rate of change). The multivariate squared-exponential kernel is given by

\begin{equation}\label{eq:se}
  \begin{aligned}
    \skernel[\text{Squared Exponential}]{\x}{\x'} =
    \alpha^2
    \exp\left(- \frac{ 1}{2\rho^2} \sum^\ds_{i = 1} \: \lvert x_{i,t} - x_{i,t}'
    \rvert^2 \right).
  \end{aligned}
\end{equation}

The squared-exponential kernel, has two hyperparameters. The marginal variance
$\alpha$ controls the overall variability of the GP at each input point and
determines how far the functions it describes can deviate from the mean
function. The lengthscale $\rho$ controls how quickly the covariance between
inputs decreases depending on their distance. A large lengthscale means that
the functions described by the GP are highly correlated even at input points
that are far away from each other, resulting in functions that show only slow
and steady variations. Contrary to this, a small lengthscale means that the
correlations decrease quickly, and the GP can vary more rapidly
\parencite{rasmussenGaussianProcessesMachine2006}\footnote{Although the kernel
  above uses the same lengthscale for all predictor dimensions, it is possible to
  assign separate lengthscales to each dimension. This can be useful when the
  function is expected to vary at different rates across predictors
  \parencite{rasmussenGaussianProcessesMachine2006}. For simplicity, we use a
  single shared lengthscale in this paper.}. The effect of the lengthscale
is therefore directly tied to the scale of the state variables.

\subsubsection{Hilbert space approximation}

Implementing the GP-SSM described above is challenging in practice due to the
strong dependencies within the model and its substantial computational
complexity \parencite{frigolaVariationalGaussianProcess2014}\footnote{In our
setting with $T-1$ input points, evaluating the likelihood of the full
multi-output GP requires inverting a $(T-1)$-by-$(T-1)$ covariance matrix.}.
These challenges can be addressed by approximating the GP using Hilbert space
methods \parencite{solinHilbertSpaceMethods2019,
svenssonComputationallyEfficientBayesian2016}. The key idea of this
approximation is to represent the GP through a finite set of basis functions,
which turns the GP prior into a Bayesian regression model that is linear in its
parameters\footnote{Although, the regression weights of this basis function
regression still depend nonlinearly on the hyperparameters of the GP.}. As a
result, the approximate GP-SSM can be estimated efficiently, with a
computational cost that scales linearly with the number of time points $T$ for a
given set of basis functions
\parencite{svenssonComputationallyEfficientBayesian2016}.

The Hilbert space approximation relies on the result that any stationary
covariance kernel $\skernel[\hpars]{\x}{\x'}$ can be represented equivalently
in terms of its spectral density $s_{\hpars}(\cdot)$. This relationship is
formalized by the \textit{Wiener-Khinchin theorem}, which states that the
covariance kernel and its spectral density are Fourier duals
\parencite{rasmussenGaussianProcessesMachine2006}. Building on this result,
\textcite{solinHilbertSpaceMethods2019} show that the pseudo-differential
covariance operator associated with $k_{\hpars}(\cdot)$ can be approximated on
any compact domain $\Omega \subset \mathbb{R}^{\ds}$, that is subject to
suitable boundary conditions, using Hilbert space methods.

More specifically, any stationary kernel can be approximated by an expansion of
the eigenfunctions of the Laplace operator on $\Omega$,

\begin{equation}\label{eq:kernel_approx}
  \begin{aligned}
    \skernel[\hpars]{\x}{\x'} & =
    \sum_{\hidx \in \mathbb{N}^\ds}
    s_{\hpars}\left(
    \sqrt{\bm{\lambda}_{\hidx}}
    \right)
    \phi_{\hidx}(\x)
    \phi_{\hidx}(\x'),                   \\
                              & \approx
    \sum_{\hidx \in \cmcal{H}}
    s_{\hpars}\left(
    \sqrt{\bm{\lambda}_{\hidx}}
    \right)
    \phi_{\hidx}(\x)
    \phi_{\hidx}(\x').
  \end{aligned}
\end{equation}

\noindent Here, $\bm{\lambda}_{\hidx}$ is a vector containing the eigenvalues of
the Laplacian, with one component for each input dimension, and
$\phi_{\hidx}(\cdot)$ is the corresponding multivariate eigenfunction. The
square root of the vector $\bm{\lambda}_{\hidx}$ is taken elementwise. For
bounded covariance kernels, the spectral densities (which determine the weight assigned to each frequency component in the basis function expansion) converge to zero at higher frequencies $\sqrt{\bm{\lambda}_{\hidx}}$. 
This means that the higher frequency terms in the sum
contribute less to the approximation of the kernel, and the possibly infinite
expansion can be further approximated by retaining only a finite number of basis
functions \parencite{solinHilbertSpaceMethods2019}. 

The eigenvalues and eigenfunctions themselves depend only on the choice of the
domain $\Omega$ and the imposed boundary conditions. A common choice for this
is a rectangular domain with Dirichlet boundary conditions
\parencite{solinHilbertSpaceMethods2019}. In that case, the multivariate
eigenpairs are given by

\begin{equation}
  \bm{\lambda}_{\hidx} =
  \bigl(
  (\tfrac{\pi h_1}{2 L_1})^2, \dots,
  (\tfrac{\pi h_{\ds}}
  {2 L_{\ds}})^2
  \bigr)^\top,
\end{equation}

\noindent and

\begin{equation}\label{eq:multivariate_eigenfunctions}
  \phi_{\hidx}(\x)  =
  \prod_{i=1}^{\ds} \phi_{h_i}(x_{i,t}) =
  \prod_{i=1}^{\ds} \frac{1}{\sqrt{L_i}} \sin \left(
  \frac{\pi h_i}{2 L_i} (x_{i,t} + L_i) \right).
\end{equation}

\noindent Here, $\hidx = (h_1, \dots, h_{\ds}) \in \cmcal{H}$ is a
multi-index over the Cartesian product of the basis function indices used in
each dimension and $L_i$ is the boundary of $\Omega$ in each dimension
\parencite{riutort-mayolPracticalHilbertSpace2023}.

Importantly, these eigenvalues and eigenfunctions are independent of
the chosen covariance kernel and its hyperparameters
\parencite{solinHilbertSpaceMethods2019}. The Hilbert space approximation
depends on the kernel exclusively through the spectral density
$s_{\hpars}(\cdot)$. For the squared-exponential kernel introduced in
Equation~\ref{eq:se}, the corresponding spectral density is

\begin{equation}\label{eq:se_spectral_density}
  s_{\hpars}(\bm{\omega}) =
  \alpha^2 (\sqrt{2\pi})^{\ds} \rho^{\ds}
  \exp \left(-\tfrac{1}{2} \rho^2 \bm{\omega}^\top \bm{\omega}
  \right).
\end{equation}

The approximation of the kernel in Equation~\ref{eq:kernel_approx} can
equivalently be written as a truncated \textit{Karhunen-Lo\`eve} expansion of
the GP \parencite{solinHilbertSpaceMethods2019}. For a multi-output
GP with a separable covariance function (Equation~\ref{eq:gp_prior})
this becomes a multivariate regression
\parencite{mukherjeeHilbertSpaceMethods2025}

\begin{equation}\label{eq:multioutput_expansion}
  f(\x) \approx
  \mathbf{B} \bm{\phi}(\x).
\end{equation}

\noindent Here $\bm{\phi}(\x)$ consists of the $H = |\cmcal{H}|$ retained basis functions and
$\mathbf{B}$ is a $\ds$-by-$H$ matrix of the corresponding regression weights.
The spectral density of the input kernel and the output covariance matrix enter
the basis function expansion in a matrix normal prior on $\mathbf{B} \sim
  \MN{\mathbf{0}}{\Sigmaout}{ \bm{S}_{\hpars}}$. The column
covariance of this matrix normal distribution is a diagonal matrix composed of the spectral densities
associated with each basis function in $\bm{\phi}(\x)$

\begin{equation}
  \bm{S}_{\hpars} =
  \operatorname{diag} \left(
  s_{\hpars} \left(
  \sqrt{\bm{\lambda}_{\hidx_1}}
  \right),
  \dots,
  s_{\hpars} \left(
  \sqrt{\bm{\lambda}_{\hidx_H}}
  \right)
  \right).
\end{equation}

Substituting the basis function representation of $f$ into the GP-SSM defined
in Equation~\ref{eq:ssm_base} yields an approximate GP-SSM that is
linear in the parameters
\parencite{svenssonComputationallyEfficientBayesian2016}

\begin{equation}\label{eq:ssm_approx_gp}
  \begin{aligned}
    \x[t+1] & \approx \mathbf{B}\bm{\phi}(\x) + \etat, \qquad
            & \etat                                           & \sim \N{\mathbf{0}}{\Sigmaeta},
    \qquad
    \mathbf{B}  \sim \MN{\mathbf{0}}{\Sigmaout}{
    \bm{S}_{\hpars}}                                                                                \\
    \y      & = \loadmat \x + \epsilont, \qquad
            & \epsilont                                       & \sim \N{\mathbf{0}}{\Sigmaepsilon}.
  \end{aligned}
\end{equation}

\section{Gibbs sampler with improved sampling efficiency}\label{estimation}

\textcite{svenssonComputationallyEfficientBayesian2016} proposed a Gibbs sampler
for drawing samples from the posterior distribution of the HSGP-SSM, combined
with a particle Gibbs sampler for sampling from the smoothing distribution of
the latent states, and a Metropolis-within-Gibbs step for targeting the
posterior of the hyperparameters. Here, the smoothing distribution denotes the
posterior distribution of the state variables conditional on the data observed
at all time points. As explained before, this algorithm shows slow mixing and
convergence when estimating the hyperparameters of the GP
(Figure~\ref{fig:orig_trace}). To address this we propose two improvements to
the algorithm, which substantially increase its sampling efficiency and
convergence. We further extend the algorithm to the CFA measurement model. 

First, we sample the hyperparameters $\hpars$ jointly with the parameters of
the dynamic model. This block sampling step improves the sampling efficiency
for the hyperparameters by reducing their dependence on the current parameter
values of the dynamic model \parencite{vandykPartiallyCollapsedGibbs2008}.
Second, we replace the bootstrap particle filter used by
\textcite{svenssonComputationallyEfficientBayesian2016} with an auxiliary
particle filter (APF). The APF would be more efficient in this setting, since
we can exploit local conjugacy within the model to construct a locally optimal
proposal distribution and auxiliary weight function
\parencite{johansenNoteAuxiliaryParticle2008}. This leads to improved mixing of
the Markov chains for the latent states and helps the sampler converge more
reliably for the same number of particles. Finally, we extend the algorithm to
structured (i.e., constrained) loading matrices $\loadmat$, which are commonly
used in structural equation modelling
\parencite{muthenBayesianStructuralEquation2012}. For this, we introduce an
extension that allows direct sampling of the free elements of the constrained
loading matrix $\loadmat$.

The technical details of the proposed Gibbs sampler can be found in
Algorithm~\ref{alg:gibbs_hilbert_gp_ssm}. The sampler was implemented in C++,
for computational efficiency, and made accessible in R \parencite{R-base} using
\texttt{Rcpp} \parencite{R-Rcpp}. The sampler can be accessed as part of the 
\href{https://github.com/Jan-Ian-Failenschmid/gpssmR}{\texttt{gpssmR}} R package and
all calculations in this paper were done using version 0.0.0.9002 
of the package. Next we will discuss the specific sampling steps of
this algorithm.

\begin{algorithm}[htbp]
  \caption{Gibbs sampler for the HSGP-SSM}
  \label{alg:gibbs_hilbert_gp_ssm}
  \begin{algorithmic}[1]
    \Require Data $\YT$ and priors on
    $\XT, \hpars, \Sigmaeta, \mathbf{B}, \Sigmaepsilon, \loadmat$

    \Ensure $K$ samples from the posterior
    $\pcond{\XT, \hpars, \Sigmaeta, \mathbf{B}, \Sigmaepsilon, \loadmat}{\YT}$

    \Statex

    \State Initialize
    $\mathbf{X}_{1:T,0}, \hpars_0, \Sigmaeta[\eta, 0], \mathbf{B}_0,
      \Sigmaepsilon[\epsilon, 0], \loadmat_0$

    \Statex

    \For{$k = 1$ \textbf{to} $K$}

    \State Sample
    $\XT[1:T,k] \sim
      \pcond{\XT}{\YT, \Sigmaeta[\eta, k-1], \mathbf{B}_{k-1},
        \Sigmaepsilon[\epsilon, k-1], \loadmat_{k-1},
        \XT[1:T, k-1]}$
    \Comment{Sampl. Sch.~\ref{alg:pgas_hilbert_gp_ssm}}

    \State Sample
    $\hpars_k \sim \pcond{\hpars}{\XT[1:T, k]}$
    \Comment{Eq.~\eqref{eq:marginalized_likelihood}}

    \State Sample
    $\Sigmaeta[\eta, k] \sim \pcond{\Sigmaeta}{\XT[1:T,k], \hpars_k}$
    \Comment{Eq.~\eqref{eq:sigmaeta_post}}

    \State Sample
    $\mathbf{B}_k \sim \pcond{\mathbf{B}
      }{\XT[1:T,k], \hpars_k, \Sigmaeta[\eta, k]}$
    \Comment{Eq.~\eqref{eq:sigmaeta_post}}

    \State Sample
    $\Sigmaepsilon[\epsilon, k] \sim
      \pcond{\Sigmaepsilon}{\YT, \XT[1:T,k], \loadmat_{k-1}}$
    \Comment{Eq.~\eqref{eq:sigmaepsilon_post}}

    \State Sample
    $\loadmat_k \sim
      \pcond{\loadmat}{\YT, \XT[1:T,k], \Sigmaepsilon[\epsilon, k]}$
    \Comment{Eq.~\eqref{eq:loadmat_post}}
    \EndFor
  \end{algorithmic}
\end{algorithm}

\subsection{Sampling Scheme 1.1: Sampling the latent states}

To sample a trajectory of the latent states $\XT[1:T,k] = (\x[1,k], \ldots,
  \x[T,k])$ from the smoothing distribution at iteration $k$ of the Gibbs
sampler (Step~3 of Algorithm~\ref{alg:gibbs_hilbert_gp_ssm}), we use an
  auxiliary particle Gibbs sampler with ancestral sampling \parencite[PGAS;
  Sampling Scheme~\ref{alg:pgas_hilbert_gp_ssm};][]{lindstenParticleGibbsAncestor2014}.

{
\floatname{algorithm}{Sampling Scheme}
\begin{algorithm}[htbp]
  \renewcommand{\thealgorithm}{1.1}
  \caption{Particle Gibbs with ancestral sampling (PGAS)}
  \label{alg:pgas_hilbert_gp_ssm}
  \begin{algorithmic}[1]
    \Require Reference trajectory $\XT[1:T, k-1]$, observations $\YT$, number of particles $N$, and current parameters $\Sigmaeta, \mathbf{B},
      \Sigmaepsilon, \loadmat$
    \Ensure New trajectory $\XT[1:T,k]$

    \Statex

    \State \alignedfor{Sample $\x[1]^{(n)} \sim \q[1]{\x[1]}$}
    {\hspace{\algorithmicindent}for $n = 1,\ldots,N-1$}

    \State Set $\x[1]^{(N)} = \x[1, k-1]$

    \State Compute and normalize weights $w^{(n)}_1$
    \Comment{Eqs.~\eqref{eq:weight_calc}-\eqref{eq:normalisation}}

    \Statex

    \For{$t = 2$ \textbf{to} $T$}

    \State \alignedfor
    {Sample $a^{(n)}_t \sim \p{a_t = j}$}
    {for $n = 1,\ldots,N-1$}
    \Comment{Eq.~\eqref{eq:ancestor_sampling}}

    \State Set $\x^{(N)} = \x[t, k-1]$

    \State Sample $a^{(N)}_t \sim \p{a_t = j}$
    \Comment{Eq.~\eqref{eq:ancestor_sampling_reference}}

    \State \alignedfor
    {Sample $\x^{(n)} \sim
        \qcond[t]{\x}{\x[t-1]^{(a^{(n)})}, \y}$}
    {for $n = 1,\ldots,N-1$}
    \Comment{Eq.~\eqref{eq:proposal_t}}

    \State Compute and normalize weights $w^{(n)}_t$
    \Comment{Eqs.~\eqref{eq:weight_calc_t}-\eqref{eq:normalisation}}

    \State \alignedfor
    {Update trajectories
      $\XT[1:t]^{(n)} =
        \bigl(
        \XT[1:t-1]^{(a^{(n)})},
        \x^{(n)}
        \bigr)$}
    {for $n = 1,\ldots,N$}

    \EndFor

    \Statex

    \State Sample index $J \sim \p{J = j}$

    \State Set $\XT[1:T,k] = \XT^{(J)}$

  \end{algorithmic}
\end{algorithm}
}

Particle Gibbs samplers are designed for Bayesian inference in nonlinear and
non-Gaussian state-space models by embedding a sequential Monte Carlo (SMC)
method, i.e., a particle filter, within a Gibbs sampling scheme
\parencite{chopinParticleGibbsSampling2015}. At each iteration of the Gibbs
sampler, a particle filter is used to generate possible trajectories of
the latent state variables from the smoothing distribution. These trajectories
are represented by an ensemble of weighted particles, where higher weights are
assigned to trajectories that explain the data better. Despite relying on a
particle approximation, the particle Gibbs sampler produces samples from the
exact posterior distribution of the latent states for any number of particles $N
\ge 2$ \parencite{andrieuParticleMarkovChain2010,
chopinParticleGibbsSampling2015}. Further increasing the number of particles
generally improves the mixing properties of the resulting Markov chain.  To
ensure that the target posterior distribution remains invariant, one particle
trajectory in the particle Gibbs sampler is kept fixed to the previously sampled
latent trajectory throughout each iteration. By convention, this reference
trajectory is assigned to the last particle $N$.

\subsubsection{Initialization}

In Step~1 of Sampling Scheme~\ref{alg:pgas_hilbert_gp_ssm}, the particle system is
initialized at time $t = 1$ by sampling states for all particles except the
last $\{\x[1]^{(n)}\}_{n=1}^{N-1}$ from an initial proposal distribution
$\q[1]{\x[1]}$. The exact choice for the proposal distribution will be
discussed later. The last particle is then set to the reference state
$\x[1]^{(N)} = \x[1, k-1]$ (Step~2). In Step~3, the unnormalized importance
weights associated with each particle are computed as follows

\begin{equation}\label{eq:weight_calc}
  w^{(n)}_1 \propto
  \frac{
    \pcond{\y[1]}{ \x[1]^{(n)}}\,
    \p{\x[1]^{(n)}}
  }{
    \q[1]{\x[1]^{(n)}}
  },
  \qquad n = 1,\ldots,N,
\end{equation}

\noindent and normalized according to

\begin{equation}\label{eq:normalisation}
  \bar{w}^{(n)}_1 =
  \frac{w^{(n)}_1}{\sum_{j=1}^N w^{(j)}_1}.
\end{equation}

\subsubsection{Propagation}

At each subsequent time point $t = 2, \ldots, T$, the particle system is
updated as follows. First, the particles are resampled implicitly by drawing a
set of ancestor indices using systematic resampling
\parencite{chopinIntroductionSequentialMonte2020}. The APF samples
the ancestor indices for the non-reference particles according to
the normalized weights from the previous time point and an auxiliary
function $v_{t-1}(\y \mid \x[t-1])$ (Step~5),

\begin{equation}\label{eq:ancestor_sampling}
  \Pbb{a^{(n)}_t = j} = \bar{w}^{(j)}_{t-1} \vcond[t-1]{\y}{\x[t-1]^{(j)}},
  \qquad n = 1,\ldots,N-1.
\end{equation}

\noindent The auxiliary function takes the observations at the next time point
into account during the resampling and allows the APF to favor ancestors that
are likely to predict $\y$ well. The optimal choice for the auxiliary function
is shown below. Afterwards, the state of the last particle is set to
the reference trajectory $\x^{(N)} = \x[t, k-1]$ (Step~6). In contrast to the
standard particle Gibbs sampler, PGAS then also samples an ancestor index for
the reference particle. This ancestor is drawn according to (Step~7),

\begin{equation}\label{eq:ancestor_sampling_reference}
  \Pbb{a^{(N)}_t = j}
  \propto
  \bar{w}^{(j)}_{t-1}\, \pcond{\x^{(N)}}{\x[t-1]^{(j)}}.
\end{equation}

In a standard particle Gibbs sampler, the history of the reference particle is
fixed to the reference trajectory \parencite{chopinParticleGibbsSampling2015}.
Consequently, any particle that has the reference particle as its ancestor
inherits the same fixed trajectory for all previous time points. Because the
probability that a particle traces back to the reference particle increases
with more time points, the ancestral paths of the particles at early time
points gradually collapse onto the reference trajectory as the algorithm
progresses. This phenomenon, often called path degeneracy, can considerably
reduce the sampling efficiency for earlier time points
\parencite{lindstenParticleGibbsAncestor2014}. PGAS prevents this issue by
probabilistically reassigning a new ancestor to the reference particle at each
time step. By allowing the history of the reference trajectory to change, PGAS
reduces path degeneracy and is able to achieve a higher sampling efficiency
across time points, even when using a relatively small number of particles
\parencite{lindstenParticleGibbsAncestor2014}.

Given the sampled ancestor indices, new states for the non-reference particles
are proposed according to a proposal distribution (Step~8),
\begin{equation}\label{eq:proposal_t}
  \x[t]^{(n)} \sim
  \qcond{\x}{\x[t-1]^{(a^{(n)})}, \y},
  \qquad n = 1,\ldots,N-1.
\end{equation}

In Step~9 the importance weights for all particles are then updated according
to

\begin{equation}\label{eq:weight_calc_t}
  w^{(n)}_t \propto
  \frac{ \pcond{\y}{\x^{(n)}} \pcond{\x^{(n)}}{\x[t-1]^{(a^{(n)})}}}{
    \vcond{\y}{\x[t-1]^{(a^{(n)})}} \qcond{\x^{(n)}}{\x[t-1]^{(a^{(n)})}, \y}
  }, \qquad n = 1,\ldots,N,
\end{equation}

\noindent and normalized.

Afterwards, the particle trajectories of each particle are updated by inheriting
the full history of their respective ancestors and appending the newly
generated states to them (Step~10),
\begin{equation}
  \XT[1:t]^{(n)} = \left( \XT[1:t-1]^{(a^{(n)})}, \x[t]^{(n)} \right),
  \qquad n = 1,\ldots,N.
\end{equation}

After propagating the particle ensemble across all time points, a complete trajectory is sampled according to the normalized importance weights at the final time point (Step~12). This trajectory is then returned as the next MCMC sample (Step~13).

\subsubsection{
  Choosing the optimal proposal distribution and auxiliary function}

Another well-known limitation of SMC methods is particle degeneracy
\parencite{chopinIntroductionSequentialMonte2020}. Particle degeneracy occurs
when, at a given time point, the observations are likely for only a small
subset of the particles. As a result, the importance weights
(Equations~\ref{eq:weight_calc} and \ref{eq:weight_calc_t}) become highly
concentrated on a few particles, while the remaining particles receive weights
close to zero. This substantially reduces the effective sample size and may
collapse the particle system.

The bootstrap proposal distribution,

\begin{equation}
  \qcond{\x}{\x[t-1], \y} = \pcond{\x}{\x[t-1]},
\end{equation}

\noindent used by \textcite{svenssonComputationallyEfficientBayesian2016}, is
prone to particle degeneracy when the observations are highly informative about
the latent states or when the predictions made by the dynamic model are
inaccurate, which is likely during early iterations of the Gibbs sampler before
the parameters have moved into regions with high posterior probability.

Particle degeneracy can be reduced by using a locally optimal proposal
distribution, which minimizes the variance of the importance weights at each
time point. This proposal corresponds to the conditional posterior
$\pcond{\x}{\x[t-1], \y}$. In the presented model, this distribution can be
derived analytically, since $\pcond{\y}{\x}$, $\pcond{\x}{\x[t-1]}$, and
$\p{\x[1]}$ are multivariate normal. The resulting proposal is given by

\begin{equation}
  \begin{aligned}
    \qcond{\x}{\x[t-1], \y} & = \pcond{\x}{\x[t-1], \y} =
    \N{\bm{\mu}_{q,t}}{\bm{\Sigma}_{q,t}},                    \\
    \text{with}\ \bm{\mu}_{q,t}          & = \bm{\mu}^* + \mathbf{K}
    \left(\y - \loadmat \bm{\mu}^* \right) ,                  \\
    \bm{\Sigma}_{q,t}       & = \bm{\Sigma}^* -
    \mathbf{K} \loadmat \bm{\Sigma}^*, \\
    \mathbf{K}              & = \bm{\Sigma}^* \loadmat^{\top}
    \left(
    \loadmat \bm{\Sigma}^* \loadmat^{\top} + \bm{\Sigma}_{\bm{\epsilon}}
    \right)^{-1}.                                             
  \end{aligned}
\end{equation}

\noindent Here, $\mathbf{K}$ denotes the Kalman gain and the quantities
$\bm{\mu}^*$ and $\bm{\Sigma}^*$ correspond to the mean and covariance of the
initial state distribution (Equation~\ref{eq:init_dist}) for $t = 1$, or of the
state transition distribution

\begin{equation}
  \pcond{\x}{\x[t-1]} = \N{\mathbf{B}\bm{\phi}(\x[t-1])}{\Sigmaeta},
\end{equation}

\noindent for making proposals at all subsequent time points.

Similarly, the optimal auxiliary function is given by

\begin{equation}
  \begin{aligned}
    \vcond{\y}{\x[t-1]} & = \pcond{\y}{\x[t-1]} =
    \N{\loadmat \mathbf{B}\bm{\phi}(\x[t-1])}{
      \loadmat \Sigmaeta \loadmat^\top + \Sigmaepsilon}.
  \end{aligned}
\end{equation}

\noindent An APF that uses both the locally optimal proposal and auxiliary
function is sometimes referred to as `perfectly adapted', as it yields
importance weights (Equation~\ref{eq:weight_calc_t}) that are equal to one and
thus have zero variance
\parencite{pittAuxiliaryVariableBased2001,pittFilteringSimulationAuxiliary1999}.
In this case, the weight computation Step~9 in
Sampling Scheme~\ref{alg:pgas_hilbert_gp_ssm} can be omitted.

On a cautionary note, this does not imply that the global efficiency of the APF
is necessarily sufficient, and it can still suffer from particle degeneracy
\parencite{chopinIntroductionSequentialMonte2020,
johansenNoteAuxiliaryParticle2008}. It is therefore still advised to carefully
check the convergence of the MCMC chains it produces. In our testing, using the
APF generally improved the convergence of the Gibbs sampler, enhanced the mixing
of the resulting Markov chains, and reduced the number of particles required by
the PGAS algorithm.

\subsection{Sampling Scheme 1.2: Sampling the parameters of the dynamic model}

In Steps~5 and 6 of Algorithm~\ref{alg:gibbs_hilbert_gp_ssm}, we sample the
parameters of the dynamic model from their conditional posterior distributions.
Following \textcite{svenssonComputationallyEfficientBayesian2016}, we set the
output covariance of the GP equal to the dynamic error covariance
$\Sigmaout = \Sigmaeta$ and assign it an
inverse-Wishart prior,

\begin{equation}
  \p{\Sigmaeta} = \IW{\vprioreta}{\Psiprioreta}.
\end{equation}

\noindent This yields a fully conjugate matrix-normal inverse-Wishart
distribution, together with the matrix normal prior on $\mathbf{B}$
\parencite[Equation~\ref{eq:ssm_approx_gp}][]{willsEstimationLinearSystems2012}.

For notational convenience, we summarize the contribution of the sampled
latent states in the following sufficient statistics:

\begin{equation}
  \begin{aligned}
    \bm{Z}      & = \XT[2:T] \XT[2:T]^\top,              \\
    \bm{\Pi}    & = \XT[2:T] \bm{\Phi}(\XT[1:T-1])^\top, \\
    \bm{\Delta} & = \bm{\Phi}(\XT[1:T-1])
    \bm{\Phi}(\XT[1:T-1])^\top,                          \\
    n           & = T - 1.
  \end{aligned}
\end{equation}

\noindent Using these sufficient statistics, the conditional posterior
distributions of the dynamic regression coefficients and error covariance
matrix can be written as

\begin{equation}\label{eq:sigmaeta_post}
  \begin{aligned}
    \pcond{\Sigmaeta}{\XT, \hpars}
                  & = \IW{\vposteta}{\Psiposteta},                          \\
    \pcond{\bm{B}}{\XT, \hpars, \Sigmaeta}
                  & = \MN{\bm{\Pi}\Sigmapostcol}{\Sigmaeta}{\Sigmapostcol}, \\
    \text{with}\ \Sigmapostcol & =
    \left(\bm{\Delta} + \bm{S}^{-1}_{\hpars} \right)^{-1},                  \\
    \vposteta     & = \vprioreta + n,                                       \\
    \Psiposteta   & = \Psiprioreta + \bm{Z}
    - \bm{\Pi} \Sigmapostcol \bm{\Pi}^\top,
  \end{aligned}
\end{equation}

\noindent and sampled directly from these conditional posterior
distributions.

\subsection{Sampling Scheme 1.3: Sampling the parameters of the measurement model}

For sampling the parameters of the measurement model in Steps~7 and 8 of
Algorithm~\ref{alg:gibbs_hilbert_gp_ssm}, it is common, but currently not
supported by the GP-SSM, to use a structured loading matrix $\loadmat$, where
certain entries have been constrained to known values. These constraints serve
two purposes. First, they allow researchers to impose a specific structure on
the measurement model (through zero constraints). Second, constraining the
loading of one observed indicator per state to one reduces the identifiability
issues by anchoring the scale of each state variable to the scale of a
corresponding indicator variable
\parencite{hayashiExaminingIdentificationIssues2006}. These indicator variables
are referred to as anchor variables. A Gibbs sampler for the CFA model can be
found in \textcite{asparouhovBayesianAnalysisUsing2023}.

The loading matrix is
vectorized and partitioned into a vector of free loadings $\bm{\lambda}$ and a
vector of fixed loadings $\bm{\lambda}^*$. The free loadings are assigned a
multivariate normal prior,

\begin{equation}
  \p{\bm{\lambda}} = \N{\mupriorlambda}{\Sigmapriorlambda},
\end{equation}

\noindent and the measurement error covariance matrix is given an inverse-Wishart prior,

\begin{equation}
  \p{\Sigmaepsilon} = \IW{\vpriorepsilon}{\Psipriorepsilon}.
\end{equation}

Because these priors are not jointly conjugate, their conditional posterior
distributions are sampled from in separate Gibbs steps. Sampling from the
conditional posterior of the measurement error covariance matrix is
straightforward by following

\begin{equation}\label{eq:sigmaepsilon_post}
  \begin{aligned}
    \pcond{\Sigmaepsilon}{\YT, \XT, \loadmat} & =
    \IW{\vpostepsilon}{\Psipostepsilon},                              \\
    \text{with}\ \vpostepsilon                 & = \vpriorepsilon + T, \\
    \Psipostepsilon                           & = \Psipriorepsilon +
    (\YT - \loadmat \XT)(\YT - \loadmat \XT)^\top.
  \end{aligned}
\end{equation}

Sampling from the conditional posterior of the measurement loadings is slightly
more involved. To this end, we define two transformations of $\x$, denoted by
$\bm{V}_t$ and $\bm{V}^*_t$. The matrix $\bm{V}_t$ contains the columns of
$\bm{I} \otimes \x^\top$ corresponding to the free loadings $\bm{\lambda}$, whereas
$\bm{V}^*_t$ contains the columns corresponding to the fixed loadings
$\bm{\lambda}^*$\footnote{
  The order of the columns of $\bm{I} \otimes \x^\top$ correspond to a row-wise
  vectorization of $\loadmat$. We therefore, first, reorder the columns to
  align with the column-wise vectorization of $\loadmat$. Afterwards, we
  extract the columns that belong to $\bm{\lambda}$ and $\bm{\lambda}^*$.
}. Using these transformations, the following sufficient statistics can be
constructed

\begin{equation}
  \begin{aligned}
    \bm{\Xi} & = \sum_{t = 1}^T \bm{V}_t^\top \Sigmaepsilon^{-1} \bm{V}_t, \\
    \bm{\xi} & = \sum_{t = 1}^T \bm{V}_t^\top \Sigmaepsilon^{-1} \left(
    \y - \bm{V}^*_t \bm{\lambda}^*
    \right).
  \end{aligned}
\end{equation}

\noindent We can then use these sufficient statistics to form the posterior distribution
of the free loadings

\begin{equation}\label{eq:loadmat_post}
  \begin{aligned}
    \pcond{\bm{\lambda}}{\YT, \XT, \Sigmaepsilon} & =
    \N{\mupostlambda}{\Sigmapostlambda},                               \\
    \text{with}\ \Sigmapostlambda                 & =
    \left( \left(\Sigmapriorlambda \right)^{-1} + \bm{\Xi} \right)^{-1}, \\
    \mupostlambda                             & =
    \Sigmapostlambda
    \left( \bm{\xi} + \left(\Sigmapriorlambda \right)^{-1}
    \mupriorlambda\right).
  \end{aligned}
\end{equation}

\subsection{Sampling Scheme 1.4: Sampling the hyperparameters of the GP}

To sample from the posterior distribution of the hyperparameters of the GP
in Steps~3 of Algorithm~\ref{alg:gibbs_hilbert_gp_ssm}, we also use a
Metropolis-within-Gibbs step. However, in contrast to
\textcite{svenssonComputationallyEfficientBayesian2016}, we block the
hyperparameters together with the parameters of the dynamic model
\parencite{vandykPartiallyCollapsedGibbs2008}, which increases the mixing of the
hyperparameter chains considerably. We therefore use the marginalized
likelihood of the hyperparameters in the Metropolis-Hastings acceptance
ratio \parencite{hassanIntroductionBayesianEconometrics2026}

\begin{equation}\label{eq:marginalized_likelihood}
  \begin{aligned}
    \pcond{\XT}{\hpars} & = \frac{
      \fdet[\frac{\ds}{2}]{\Sigmapostcol}
      \fdet[\frac{\vprioreta}{2}]{\Psiprioreta}
      \fgamma[\frac{\vposteta}{2}]{\ds}
    }{
      \pi^{\frac{\ds n}{2}}
      \fdet[\frac{\ds}{2}]{\bm{S}_{\hpars}}
      \fdet[\frac{\vposteta}{2}]{\Psiposteta}
      \fgamma[\frac{\vprioreta}{2}]{\ds}
    }.
  \end{aligned}
\end{equation}

\noindent Here $\Gamma(\cdot)$ is the multivariate Gamma function. The prior for
the hyperparameters can be freely specified as Metropolis-Hastings does not
require conjugacy. Since the number of hyperparameters is relatively small, we
use a random-walk proposal for the Metropolis-Hastings step and adaptively tune
the proposal distribution to an average acceptance rate of .234
\parencite{haarioAdaptiveMetropolisAlgorithm2001,hoffmanNoUTurnSamplerAdaptively2014}.

\subsection{Tuning the Hilbert space approximation}\label{hsgp_tuning}

The Hilbert space approximation has several tuning parameters that need to be
selected carefully for accurate inference. In particular, these are the number
of basis functions $J_i$ and the location of the domain boundary $L_i$ of
$\bm{\Omega}$ in each dimension $i \in \{1, \dots, \ds\}$.
\textcite{riutort-mayolPracticalHilbertSpace2023} describe a procedure for
finding appropriate values for the tuning parameters based on the location of
the input points and the GPs lengthscale.
They recommend defining the boundary of the approximation domain as

\begin{equation}
  L_i = c_i L^*_i,
\end{equation}

\noindent  where $c_i$ is a boundary expansion factor and $L^*_i$ corresponds,
in our case, to the maximum absolute value of the $i$-th state.
Since the states are unobserved, we instead start by setting $L^*_i$
to the maximum absolute value of each corresponding anchor variable.

Once $L^*_i$ is specified, the number of basis functions $J_i$ and the boundary
expansion factor $c_i$ can be set as functions of the lengthscale $\rho$. In
general, shorter lengthscales require more basis functions, whereas longer
lengthscales require a wider boundary expansion factor. For the squared
exponential kernel, \textcite{riutort-mayolPracticalHilbertSpace2023} recommend

\begin{equation}\label{eq:tuning_pars}
  \begin{aligned}
    c_i \geq 3.2 \, \frac{\rho}{L^*_i} \quad & \& \quad c_i \geq 1.2,              \\
    J_i                                      & = 1.75 \, \frac{c_i L^*_i}{\rho}. \\
  \end{aligned}
\end{equation}

\noindent  Because the lengthscale is also generally unknown, the tuning
parameters of the Hilbert space approximation must be updated iteratively. This
can be done by fitting the GP-SSM with an initial set of tuning parameters and
then adjusting them according to the estimated lengthscale \parencite[see
Section 4.5.1 in][]{riutort-mayolPracticalHilbertSpace2023}. During each
update, $L^*_i$ can additionally be set to the maximum absolute value of each
estimated state in the previous iteration.

\subsection{Initializing the sampler}

\textcite{svenssonComputationallyEfficientBayesian2016} showed that their
algorithm converges asymptotically to the posterior distribution of the
HSGP-SSM. In practice, however, it is known that data augmentation algorithms
(i.e., algorithms that sample latent variables alongside the model parameters)
can converge slowly, particularly when the observations are weakly informative
about the state variables
\parencite{yuCenterNotCenter2011,papaspiliopoulosGeneralFrameworkParametrization2007}.
In addition to this, poor initial values for the model parameters may lead to
particle degeneracy in Sampling Scheme~\ref{alg:pgas_hilbert_gp_ssm}, which may
further slow down the convergence. For these reasons, it is important to use
appropriate starting values for the Gibbs sampler.

To obtain suitable starting values, we use an initial guess for the latent
states and then sample the remaining parameters from their respective posterior
distributions conditional on this guess. One possible strategy for this is to
fit a simpler (e.g., linear) model and use its fitted states as initial values
\parencite{svenssonFlexibleStateSpace2017,paduartIdentificationNonlinearSystems2010}.
For this version of the HSGP-SSM an even simpler option is available. Since,
each anchor variable (i.e., the variables with a fixed loading of one for one
latent state) is a noisy indicator for one state and is guaranteed to be on the
same scale as the state, we can use the observed anchor variables as starting
values for their respective states. During the updating of the HSGP tuning
parameters, this can be further refined and replaced with a sample from the
smoothing distribution of the model fitted at the previous iteration. 

\section{Simulation-Based Calibration}\label{sbc}

Since we propose a new software implementation for an adjusted sampler, it is
important to evaluate whether the model and sampler are well calibrated. In
this context, calibration means that for a given prior distribution, the
sampler is able to produce draws from the correct posterior distribution. A
probabilistically principled way to evaluate a Bayesian model is
simulation-based calibration \parencite[SBC;][
]{modrakSimulationBasedCalibrationChecking2022,taltsValidatingBayesianInference2020}.

To conduct the SBC, we first specify informative prior distributions for all
parameters. We then draw sets of parameters from the joint
prior and simulate a data set for each parameter set. Afterwards, we fit the
HSGP-SSM to each data set and compute the rank of the true parameter value
among the posterior samples. If the model and sampler are well calibrated,
these ranks should follow a uniform distribution. This also directly implies
that the posterior credible intervals have their nominal coverage probabilities
\parencite{modrakSimulationBasedCalibrationChecking2022}. This procedure can
easily be performed using the \texttt{SBC} R package \parencite{R-SBC}.

SBC relies on a self-consistency property of Bayesian models. Specifically, if
parameters are drawn from the prior, data are generated conditional on these
parameters, and posterior samples are then obtained given the data, the
resulting prior and posterior draws are, conditional on the data, identically
distributed according to the posterior
\parencite{modrakSimulationBasedCalibrationChecking2022}. This property follows
directly from Bayes' theorem, where the posterior distribution is obtained by
conditioning the prior on the observed data, and is closely related to the fact
that the data-averaged posterior of a Bayesian model equals its prior.
Consequently, if the model and sampler are correctly implemented, parameter
values drawn from the prior are uniformly distributed within the corresponding
posterior distribution.

By generating data from the exact model and fitting the approximate
HSGP-SSM, we additionally assess the consistency of the Hilbert space
approximation. The calibration of multivariate Hilbert space approximate GPs
has been successfully established for other latent variable models
\parencite{mukherjeeHilbertSpaceMethods2025}, but to our knowledge, it has not
yet been evaluated in the context of GP-SSMs.

\subsection{Data generation}

We generated parameters and data from GP-SSMs with one and two latent states.
For each state, we used four indicator variables. The loading of the first
indicator variable for each state was fixed to one. Additionally, in the
bivariate model, all cross-loadings were fixed to zero. In both models, the GP
used a single set of hyperparameters with a squared exponential kernel.

The priors were specified as follows

\begin{equation}\label{eq:priors}
  \begin{aligned}
    \x[1]         & \sim \N{\bm{0}}{5 \, \identity[\ds]},            \qquad
    & \alpha        & \sim \IG{12}{45},                                \\
    \rho          & \sim \IG{12}{\sqrt{\ds} \, 45},                  \qquad
    & \Sigmaeta     & \sim \IW{24 + \ds - 1}{ 90 \, \identity[\ds] },  \\
    \bm{\lambda}  & \sim \N{\bm{0}}{2 \, \identity[d_\lambda]},      \qquad
    & \Sigmaepsilon & \sim \IW{4 + \dobs - 1}{70 \, \identity[\dobs]}.
  \end{aligned}
\end{equation}

\begin{figure}[!ht]
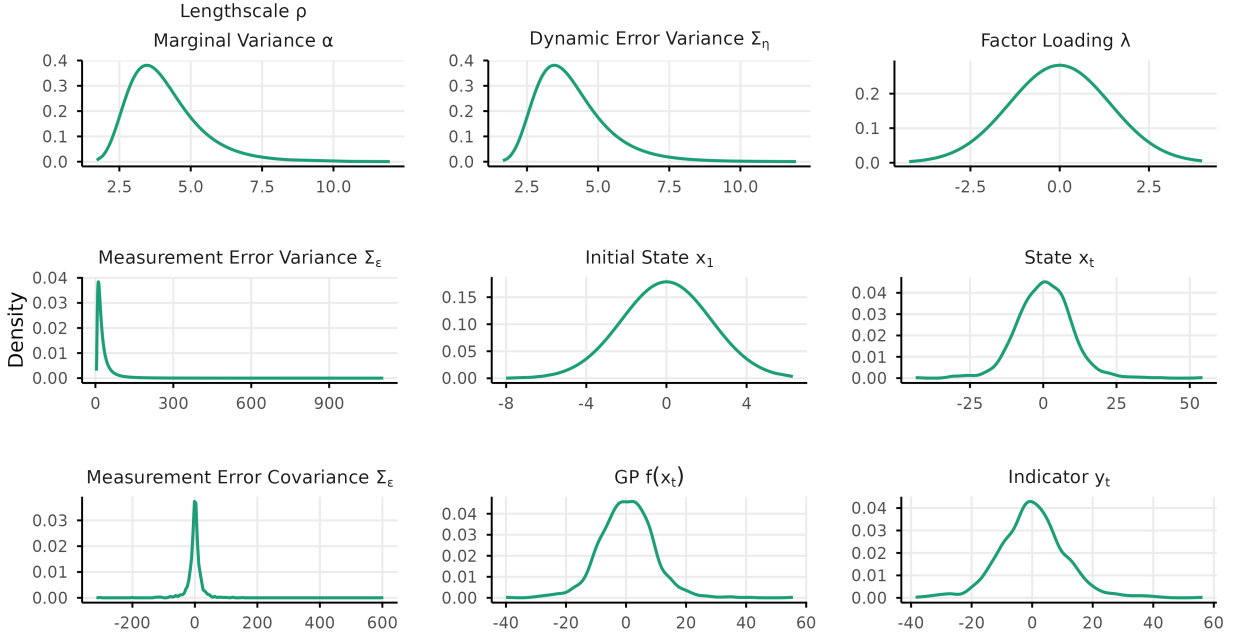

  \caption{Prior distributions}
  \fitfigure{prior_density.png}
  \figurenote{This figure shows the prior distributions in Equation~\ref
  {eq:priors} for the single state model. For parameters that occur multiple
  times in the model and have identical priors, such as $\x$, a single example
  is shown. For parameters or derived quantities that have a distribution without a simple analytic expression kernel density estimates are depicted
  instead.}
  \label{fig:priors}
\end{figure}

\noindent Here, $IG$ denotes the inverse-Gamma distribution, and $d_\lambda$ is
the number of unconstrained loading parameters, which equals three in the
univariate model and six in the bivariate model. These priors are illustrated
in Figure~\ref{fig:priors}.

These priors were chosen to balance simulating a wide range of dynamic systems
and generating parameter values that are suitable for testing the Gibbs
sampler. The priors for the GP hyperparameters were selected to generate
lengthscales that are neither much smaller than the spacing of the data points
nor much larger than the overall range of the data, while still allowing for a
large range of possible dynamic functions (Figure~\ref{fig:gp_priors}),
including fairly stable and fairly wiggly functions. The priors for the free
loadings were chosen to be, on average, on a comparable scale to the fixed
loading of the anchor variables (i.e., one). Finally, the prior for the
measurement error covariance was selected to yield an average $R^2$ of
approximately 0.5 for each indicator, while covering the full range of possible
$R^2$ values. This allows us to evaluate the sampler across settings in which
the data are weakly to strongly informative about the latent states, while
ensuring that, on average, the data remain sufficiently informative about the
hyperparameters so that their posterior distributions are not entirely
dominated by their priors.

\begin{figure}[!ht]
  \caption{GP prior and state trajectory draws}
  \fitfigure{prior_trajectories.png}
  \figurenote{This figure shows six independently drawn dynamic functions
    $f(\x)$ from the GP prior used for the single state models. The bottom
    panels show the state trajectories generated by each of the dynamic
    functions.}
  \label{fig:gp_priors}
\end{figure}

\noindent Using these priors, we generated 200 parameter sets and simulated
data for 200 time points for each parameter set.

\subsection{Model fitting}

Following the SBC procedure, the same priors were used to fit the HSGP-SSM.
The Hilbert space approximation was tuned according to the guidelines by
\textcite{riutort-mayolPracticalHilbertSpace2023}, using the true simulated
states and lengthscales for determining the domain boundary and the number of
basis functions. This allowed us to tune the approximation reliably across many
datasets without refitting the model multiple times per dataset. For the same
reason, we used slightly more conservative tuning parameters by setting the
minimum boundary expansion factor to 1.6 and adding five additional basis
functions to Equation~\ref{eq:tuning_pars}.

The PGAS was run with 20 particles for the univariate model and 28 particles
for the bivariate model. For the hyperparameters, we performed three
Metropolis--Hastings updates within each Gibbs iteration.

The sampler was run for 70{,}000 iterations with four independent chains. The
first 10{,}000 iterations were discarded as burn-in, and the remaining draws
were thinned by a factor of ten. The convergence during the simulation was
assessed using the scale reduction factor $\hat{R} < 1.01$ criterion
\parencite{vehtariRankNormalizationFoldingLocalization2021}. If the sampler did
not converge for a specific dataset, it was restarted up to two times. Because
SBC requires nearly independent draws for the rank calculations, we further
thinned the posterior draws by a factor of 20 for this step
\parencite{sailynojaGraphicalTestDiscrete2022}. For all other analyzes, the
full sample of 24{,}000 draws was used.

This procedure was evaluated on ten test datasets per condition and yielded,
in all cases, an effective sample size larger than 100 times the number of
chains\footnote{The models with one state were fit on an Intel i7-13700k 3.40 GHz CPU and the models with two states were fit on the Snellius supercomputer.}.

\subsection{Calibration Results}

For presenting the results of the SBC, we grouped together all matrix-valued
parameters that are treated equivalently within the sampler.

\subsubsection{Convergence}

To assess the convergence of the sampler after the simulation we used the
$\hat{R}$, the bulk effective sample size (Bulk ESS), and the tail effective
sample size (Tail ESS). Ideally, $\hat{R}$ should not exceed the threshold of
$1.01$, and the ESS should be larger than $400$ when running four independent
chains \parencite{vehtariRankNormalizationFoldingLocalization2021}. This
effective sample size is required to reliably estimate $\hat{R}$ as well as any
posterior quantities of interest. The Bulk ESS provides information about how
reliable estimates of central tendency, such as the posterior mean or mode, are,
whereas the Tail ESS provides information about how reliable the boundaries of
posterior credible intervals are.

\begin{figure}[!ht]
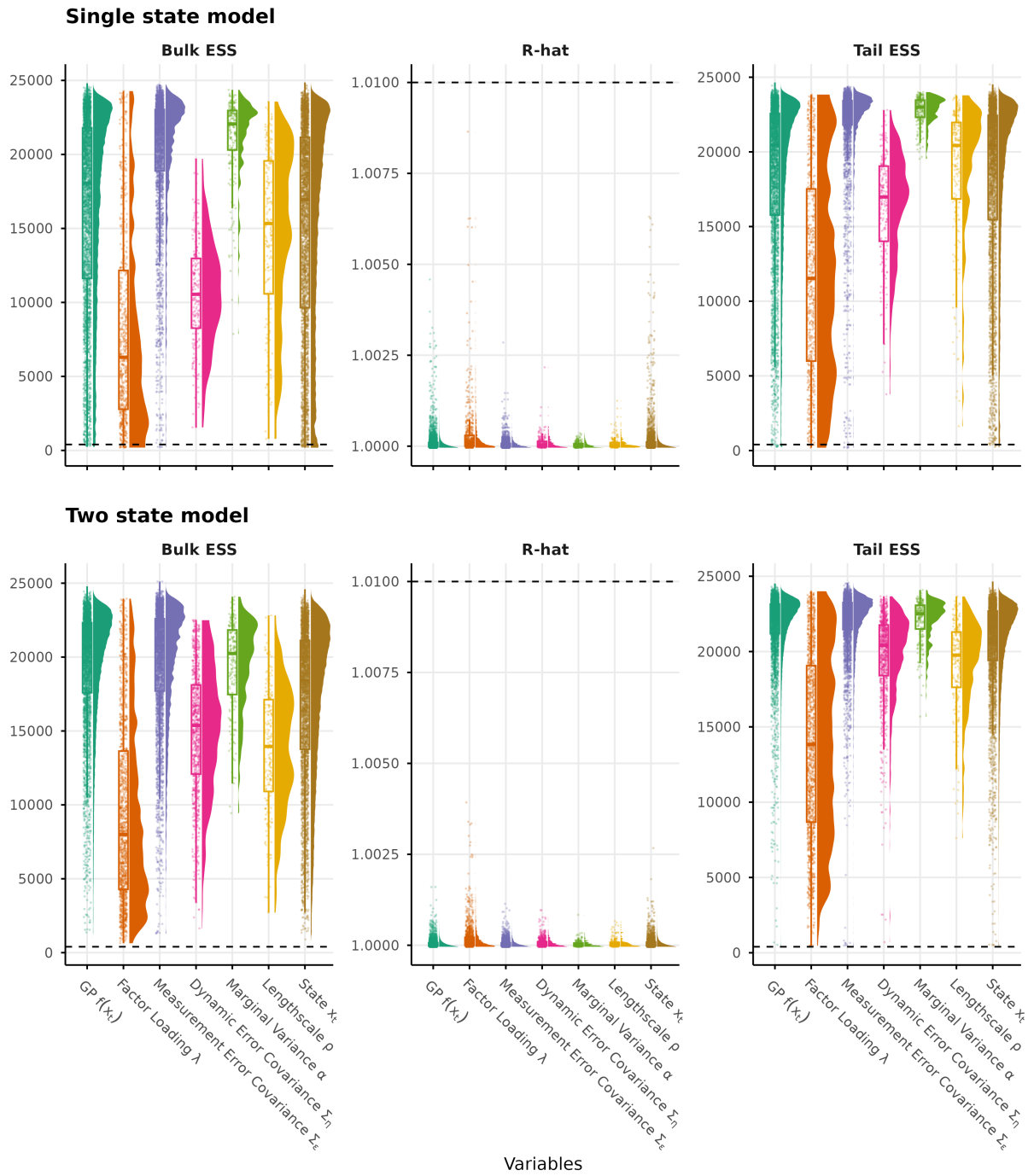

  \caption{Convergence statistics}
  \fitfigure{convergence.png}
  \figurenote{This figure shows the Bulk ESS, Tail ESS, and $\hat{R}$ for
    estimating the HSGP-SSMs with one and two latent states during the SBC. The
    horizontal lines indicate the convergence thresholds for each metric. Note
    that for the states and the GP, only 10{,}000 data points are displayed.}
  \label{fig:convergence}
\end{figure}

Figure~\ref{fig:convergence} shows the convergence statistics for all models
fitted during the SBC. All models with a single state converged according to the
$\hat{R}$ criterion. For a very few number of runs, either the Bulk ESS or Tail
ESS was below the cutoff of $400$ for some parameters, making the $\hat{R}$
estimate unreliable. This particularly affects the latent states, the GP
estimates, and the factor loadings. In total, $0.58\%$ of the parameters had a
Bulk ESS and $0.22\%$ had a Tail ESS below the threshold. Because the proportion
of parameters with a low ESS is relatively small we included these models in the
analysis. All models with two latent states converged according to all metrics.
The models with two latent states likely converge better because the two states
can inform each other and provide more information about the hyperparameters.
Additionally, the data for the two-state model had a slightly higher
signal-to-noise ratio.

\subsubsection{Calibration Results}

\begin{figure}[!ht]
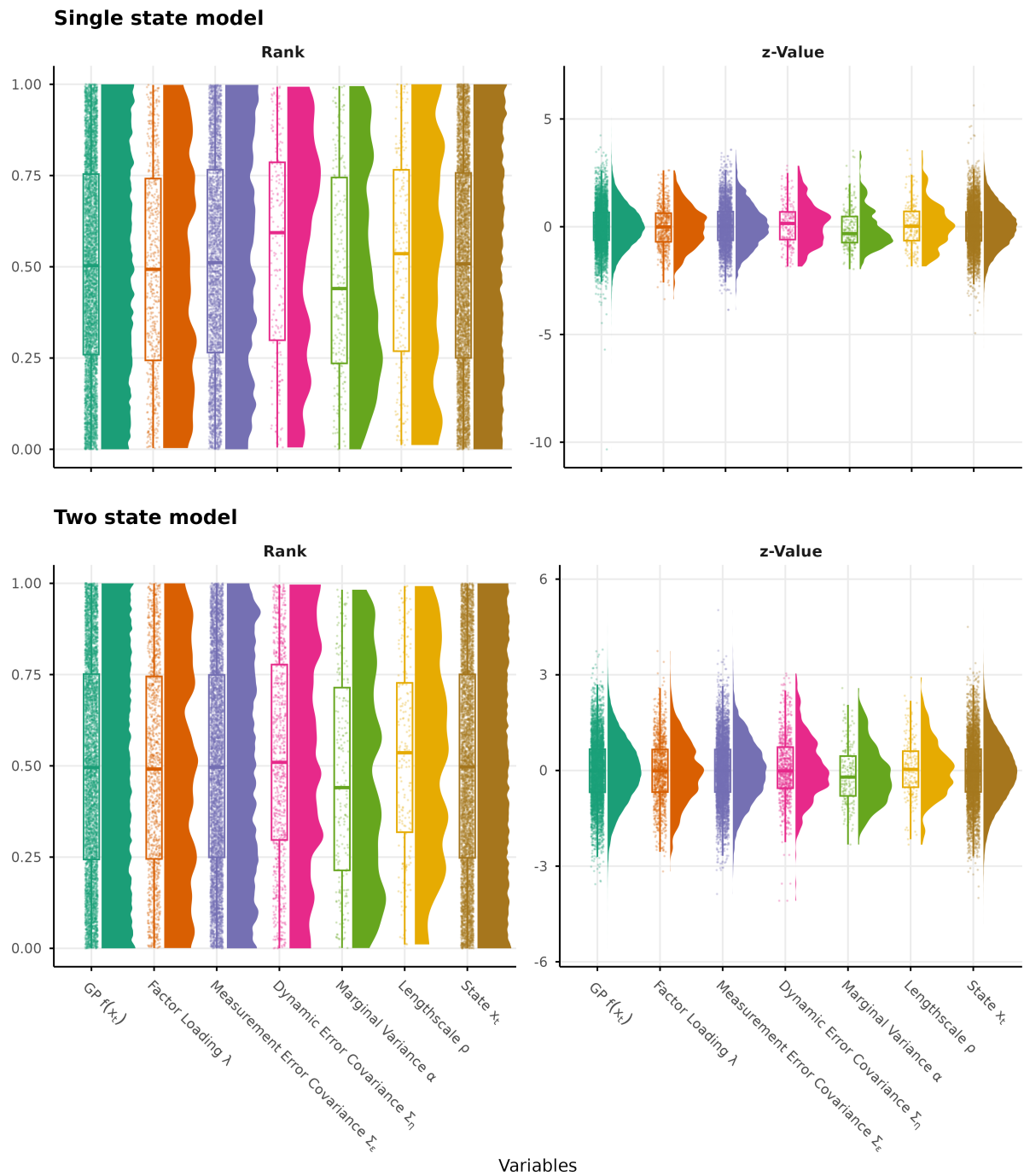

  \caption{Prior parameter ranks and $z$-scores}
  \fitfigure{raw_ranks.png}
  \figurenote{This figure shows the ranks of the simulated prior parameters
    among the posterior samples and their distributions. Additionally, it shows
    the $z$-values of the simulated parameters relative to the posterior mean and
    standard deviation. For the states and the GP, only 10{,}000 data points are
    displayed.}
  \label{fig:raw_ranks}
\end{figure}

The main outcome of the SBC is whether the ranks of the prior parameters are
uniformly distributed among the posterior draws for each dataset.
Figure~\ref{fig:raw_ranks} shows the normalized ranks of all prior parameters
and the estimated distributions of the ranks. Additionally, it shows the
$z$-scores of the prior parameters relative to the posterior means and standard
deviations.

\begin{figure}[!ht]
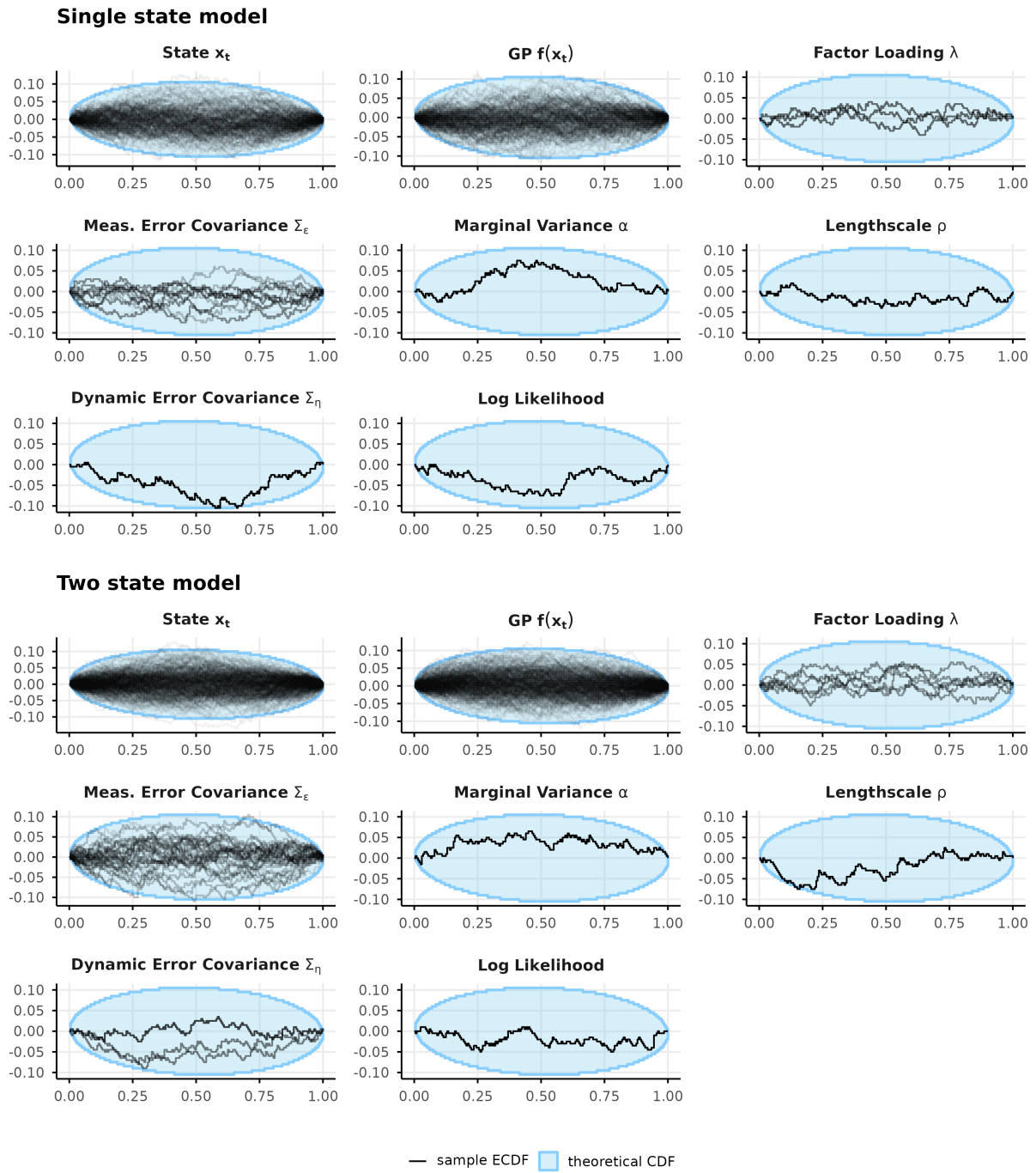

  \caption{ECDF difference}
  \fitfigure{ecdf_diff.png}
  \figurenote{This figure shows the difference between the ECDF of the ranks
    for each parameter and a theoretical uniform CDF. The blue area represents a
    $95\%$ confidence region for the largest deviation from the uniform CDF. If
    the sampler and model are well calibrated, the difference between the sample
    ECDF and the uniform CDF should remain within this region for $95\%$ of the
    parameters.}
  \label{fig:ecdf_diff}
\end{figure}

To assess whether the distribution of any parameter follows a
uniform distribution, we consider their empirical cumulative distribution
function \parencite[ECDF;][]{sailynojaGraphicalTestDiscrete2022}.
Figure~\ref{fig:ecdf_diff} shows the differences between the observed ECDF of
each parameter and a theoretical uniform CDF. The blue area corresponds to a
$95\%$ confidence region for the largest deviation from the uniform CDF. We
would therefore expect the ECDFs of $95\%$ of the parameters to remain within
this region if the sampler is well calibrated. Based on these graphical tests the sampler and the HSGP-SSM seem to be
well-calibrated\footnote{It is, however, noteworthy that in earlier tests,
particularly before adopting the APF proposal, the dynamic error variance tended
to be underestimated when too few particles (i.e., 20) were used in the PGAS and
the resulting state estimates had a low ESS at some time points. Since
Figure~\ref{fig:ecdf_diff} still shows a slight underestimation of the dynamic
error variance, we reran the simulation three additional times using different
random seeds. The results of these simulations can be found in
Appendix~\ref{app:A} and do not indicate any systematic underestimation.}.

\subsubsection{Hyperparameter and dynamic error covariance inference}

\begin{figure}[!ht]
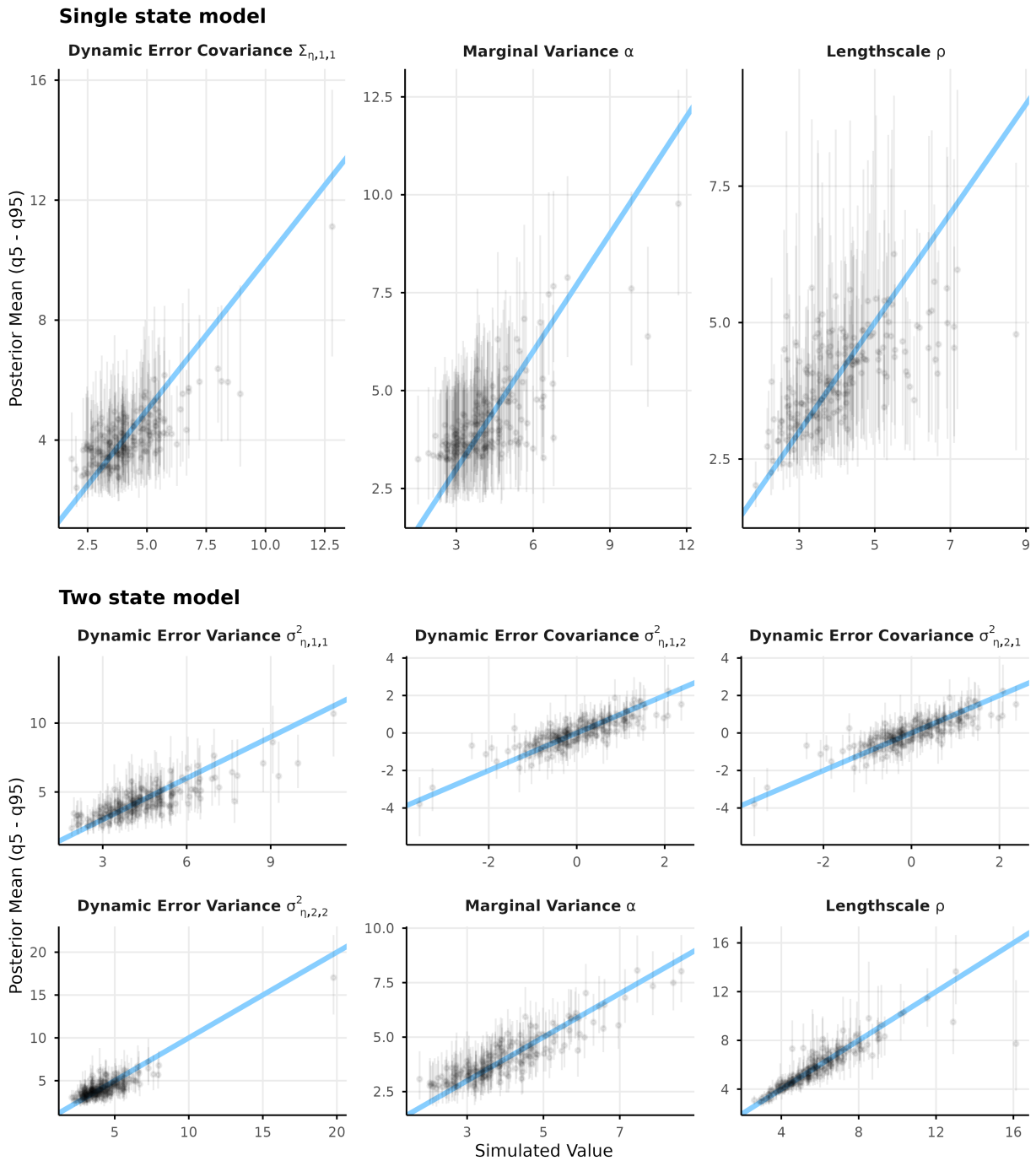

  \caption{Estimated versus simulated parameters}
  \fitfigure{sim_estimated.png}
  \figurenote{This figure shows the estimated posterior parameter means and
    $95\%$ quantile intervals for the hyperparameters and dynamic error
    covariance against their simulated values. The blue line indicates perfect
    correspondence between the simulated and estimated values.}
  \label{fig:sim_estimated}
\end{figure}

Because the hyperparameters and the dynamic error covariance were estimated with
the least precision in the SBC, we examined their inference in more detail.
Figure~\ref{fig:sim_estimated} compares the estimated values for these
parameters to their corresponding simulated values. The blue line indicates a
perfect correspondence between the simulated and estimated values. For all
parameters, it can be seen that the estimates are not perfectly centered around
this line, particularly in the single-state model. This is not uncommon in
Bayesian inference and does not mean that the posterior estimates are invalid.
Instead, it indicates that the posterior distributions of these parameters are
not entirely dominated by the likelihood and are therefore still affected by the
chosen prior. This aligns with the inference being more accurate for the model
with two states, as more information is available about the hyperparameters.

Figure~\ref{fig:sim_estimated} additionally reveals some interesting behaviors
of the posterior updating of the hyperparameters, particularly for the
single-state model. The marginal variance parameter $\alpha$ determines the
spread of the GP prior. Even for large simulated values of $\alpha$, functions
drawn from this prior can be close to zero. In that case, the posterior
distribution of $\alpha$ mainly contains values that are smaller than the
simulated value. For a similar reason, the posterior distribution rarely covers
values below the region with the highest prior mass, leading to an upward bias
for small values of $\alpha$. The lengthscale $\rho$ determines how rapidly the
functions described by the GP prior vary. Smaller lengthscales lead to a higher
signal-to-noise ratio in the GP-SSM and are therefore estimated more accurately,
whereas larger values of $\rho$ lead to posterior distributions that are more
strongly influenced by their prior.

We additionally obtained samples from the data-averaged posterior for these
parameters by subsampling 200 draws from the samples of the posterior for each
dataset and combining them. Due to the self-consistency of Bayesian models, the
data-averaged posterior and the prior distribution should be equal for each
parameter. Figure~\ref{fig:data_averaged_posterior} shows that the samples from
both distributions align closely. 

\begin{figure}[!ht]
  \caption{Data-averaged posteriors for the hyperparameters and dynamic error covariance}
  \fitfigure{data_averaged_posterior.png}
  \fitfigure{data_averaged_posterior_biv.png}
  \figurenote{This figure shows kernel density estimates of samples from the
    data-averaged posterior and the corresponding prior distributions and
    bootstrap uncertainty intervals.}
  \label{fig:data_averaged_posterior}
\end{figure}

\section{Empirical Examples}\label{emp_example}

In this section, we demonstrate the GP-SSM and the proposed Gibbs sampler
on two empirical datasets. First, we illustrate how the GP-SSM can be used
to learn the predator-prey system. Next, we apply the model to fMRI data
to demonstrate how the output of the GP-SSM can be interpreted in practice. In
order to facilitate model convergence and allow the use of comparable prior
distributions, all data presented in this section were standardized prior to
the analysis.

\subsection{Learning the predator-prey system}

The classical dataset to illustrate the predator-prey dynamics concerns the
population sizes of Canadian lynx and snowshoe hares. As indicators of their
respective population sizes, the dataset contains the number of pelts, in
thousands, purchased by the Hudson's Bay Company between 1845 and 1935. Although
these indicators are fairly indirect measures of the underlying population
sizes, the data provide a well-known example of the cyclic relationship between
the lynx and hare populations. For a detailed and critical discussion of this
dataset, see \textcite{murrayMathematicalBiology1993}.

Because the data consist of population counts, the CFA measurement model (i.e.,
the linear measurement model with additive Gaussian noise) is misspecified and a
measurement model that has been specifically designed for count data would be
more appropriate. However, the goal of this example is to illustrate the
proposed HSGP-SSM, and we therefore continue using the CFA measurement model.

Figure~\ref{fig:lv_data} shows the standardized time series for both
populations. In this figure you can clearly see the characteristic lagged 
cyclic relationship between both species. Both species oscillate 
between periods of small and large population sizes, with changes in the 
predator population lagging behind changes in the prey population 
\parencite{murrayMathematicalBiology1993}.

\begin{figure}[!ht]
  \caption{Standardized predator-prey population sizes}
  \fitfigure{lv_data.png}
  \figurenote{This figure shows the standardized numbers of Canadian lynx and
  snowshoe hare pelts purchased by the Hudson's Bay Company between 1845 and
  1935}
  \label{fig:lv_data}
\end{figure}

Since each population size is measured by exactly one indicator, we use a
diagonal loading matrix with one loading fixed to one for each species and
loadings fixed to zero across species. Hence, there are no free parameters 
in the loading matrix. To fit the model, we use the following prior 
distributions:

\begin{equation}\label{eq:lv_priors}
  \begin{aligned}
    \x[1]         & \sim \N{\bm{0}}{\identity[2]},            \qquad
    & \alpha        & \sim \IG{1}{1},                                \\
    \rho          & \sim \IG{1}{1},                  \qquad
    & \Sigmaeta     & \sim \IW{7}{\identity[2]},  \\
    \Sigmaepsilon & \sim \IW{7}{\identity[2]}.
  \end{aligned}
\end{equation}

\noindent We ran the sampler with 20 particles and three Metropolis-Hastings
updates within each Gibbs iteration. Four chains were run for 30{,}000
iterations and the first 10{,}000 iterations were discarded as warm-up. We
thinned the remaining samples by a factor of ten, resulting in a final posterior
sample of 8{,}000 draws. The HSGP approximation parameters were tuned
iteratively according to Section~\ref{hsgp_tuning} and Phase A of the tuning
procedure proposed  by \textcite{riutort-mayolPracticalHilbertSpace2023}.
The chains converged during each tuning iteration, according to the same 
metrics used in the simulation study, and the final model 
used 11 basis functions in each dimension. 

Prior and posterior predictive checks for the model can be found in
Appendix~\ref{app:B}. Figure~\ref{fig:lv_prediction} shows the posterior
predictive distribution for both species plotted as a time series. The black
line corresponds to the posterior predictive mean and the lighter shaded area to
its 95\% credible interval. The figure also shows the smoothing distribution of
both latent population sizes. Here, the dotted line represents the posterior
mean and the darker shaded area the corresponding 95\% credible interval. It can
be seen that the posterior predictive distribution reproduces the main
oscillatory patterns in the data well, which indicates that the HSGP-SSM
sufficiently captured the underlying population dynamics.

\begin{figure}[!ht]
  \caption{Posterior predictive and smoothing distribution}
  \fitfigure{lv_prediction.png}
  \figurenote{This figure shows the posterior predictive distribution and the
  smoothing distribution for the lynx and hare population sizes. The black line
  shows the posterior predictive mean and the lighter shaded area its 95\%
  credible interval. The dotted line shows the posterior mean of the smoothing
  distribution and the darker shaded area its 95\% credible interval. Because
  both indicators have fixed loadings of one on their respective states, the
  posterior predictive distribution corresponds to the smoothing distribution
  with additional measurement noise.}
  \label{fig:lv_prediction}
\end{figure}

Table~\ref{tab:lv_post_pars} contains the posterior summaries for all model
parameters. The posterior mean of the lengthscale parameter is $2.94$, which is
relatively large compared to the scale of the standardized data. As a result, we
would expect the posterior GP to only vary gradually across the state space. The
posterior estimates further provide evidence that the dynamic errors that are
affecting both species are correlated. This can happen when both latent
population sizes are affected by the same external factors, such as fur trappers
hunting both species \parencite{murrayMathematicalBiology1993}.

\begin{table}[!ht]
  \caption{Posterior parameter summaries from the predator-prey model}
\centering
\begin{tabular}{rlrrrrrrr}
  \hline
 & Variable & Mean & SD & Q5 & Q95 & R-hat & ESS Bulk & ESS TAIL \\ 
  \hline
1 & alpha & 5.99 & 3.09 & 3.03 & 11.08 & 1.00 & 6519.20 & 7566.72 \\ 
  2 & rho & 2.94 & 0.95 & 1.68 & 4.76 & 1.00 & 5989.62 & 6104.35 \\ 
  3 & Sigma\_eta[1,1] & 0.08 & 0.03 & 0.05 & 0.13 & 1.00 & 5465.99 & 6548.64 \\ 
  4 & Sigma\_eta[1,2] & 0.04 & 0.02 & 0.01 & 0.07 & 1.00 & 5694.14 & 7029.59 \\ 
  5 & Sigma\_eta[2,2] & 0.10 & 0.03 & 0.06 & 0.16 & 1.00 & 4203.72 & 6069.28 \\ 
  6 & Sigma\_epsilon[1,1] & 0.05 & 0.02 & 0.03 & 0.09 & 1.00 & 4661.55 & 5458.34 \\ 
  7 & Sigma\_epsilon[1,2] & -0.04 & 0.04 & -0.12 & 0.03 & 1.00 & 2923.04 & 4925.81 \\ 
  8 & Sigma\_epsilon[2,2] & 0.55 & 0.11 & 0.39 & 0.73 & 1.00 & 4780.56 & 6093.73 \\ 
  9 & log\_lik & -82.66 & 11.99 & -102.62 & -62.95 & 1.00 & 6738.86 & 7164.31 \\ 
   \hline
\end{tabular}

  \label{tab:lv_post_pars}
\end{table}

Figure~\ref{fig:lv_posterior_gp} shows the posterior GP. Because the GP has two
input dimensions (the current lynx and hare population sizes) and two output
dimensions (their values at the next time point), the posterior is displayed as
two separate prediction surfaces. The left panels show the predicted next lynx
population size as a function of the current lynx (i.e., $x_1$) and hare (i.e.,
$x_2$) populations, whereas the right panels show the corresponding predictions
for the next hare population size. The lower panels additionally include samples
from the smoothing distribution, highlighting the regions of the state space
that are supported by the observed data.

As expected based on the inferred lengthscale, the posterior GP varies smoothly
and appears close to linear over the range informed by the observed population
sizes. The posterior GP shows that both species affect each other, such that a
larger hare population is associated with an increase in the subsequent lynx
population, whereas a larger lynx population is associated with a decrease in
the subsequent hare population. Furthermore, the effect of each species on its
own future population size depends on the abundance of the other species,
which indicates an interaction effect between the two populations.

\begin{figure}[!ht]
  \caption{Posterior Gaussian process surface}
  \fitfigure{lv_posterior_gp.png}
  \figurenote{This figure shows the posterior GP for the predator-prey dynamics.
  Here, $x_1$ corresponds to the latent Canadian lynx population size and $x_2$
  to the latent snowshoe hare population size. Since the dynamic function has
  two inputs and two outputs, the left panels show the posterior GP for
  predicting the next lynx population size, whereas the right panels show the
  posterior GP for predicting the next hare population size. The dark purple
  surface corresponds to the posterior mean and the lighter surfaces show the
  posterior 95\% credible interval. The lower panels show the same posterior GP
  together with 100 samples from the smoothing distribution. Note, that the
  different posterior GP surfaces are not depicted using the same orientation or
  scale, as each surface is shown from the perspective that best illustrates its
  features.}
  \label{fig:lv_posterior_gp}
\end{figure}

\begin{figure}[!ht]
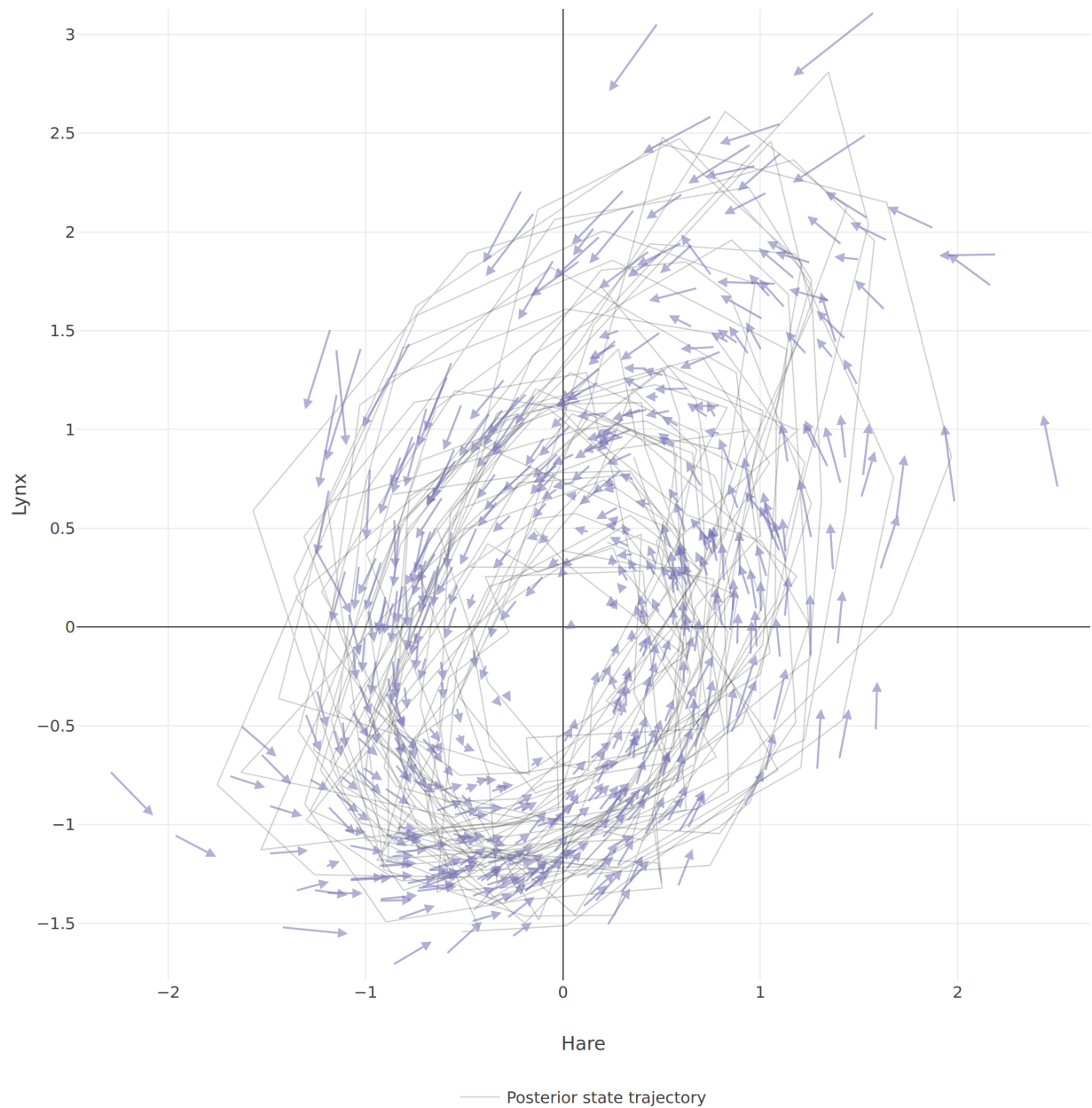

  \caption{Empirical vector field}
  \fitfigure{lv_vector_field.png}
  \figurenote{This figure shows the empirical vector field learned by the
  GP-SSM. Each arrow starts at a sample from the smoothing distribution and
  points towards the value predicted by the posterior GP for the next time point
  at 25\% scale. Overlaid are five traces from the smoothing distribution. This
  plot illustrates how the learned dynamic system changes and evolves from 
  different starting points.}
  \label{fig:lv_vector_field}
\end{figure}

Lastly, Figure~\ref{fig:lv_vector_field} shows an empirical vector field that
corresponds to the learned dynamics in phase space. Each arrow starts at a
sample from the smoothing distribution and points towards the state predicted by
the posterior GP for the next time point. The resulting vector field clearly
shows the characteristic cyclic pattern of the predator-prey relationship. It
also reproduces the characteristic phase lag between the two species. This
means that increases in the hare population are followed by increases in the
lynx population, which subsequently causes a decline in the hare population
leading eventually to a decline in the lynx population. This sequential
relationship generates the cyclic trajectory visible in the vector field and
suggests that the model was able to capture the dynamic relationship between the
two species sufficiently.

\subsection{Learning a bistable dynamic system from fMRI data}

For the second example, we use fMRI data from the first participant in
\textcite{antogniniIsofluraneAnesthesiaBlunts1997}. In this study, brain
activity was recorded every two seconds for 256 seconds while a tactile stimulus
was repeatedly applied to the participant's right palm for 32 seconds and then
removed for 32 seconds. The measurements were obtained bilaterally from the
primary and secondary somatosensory cortex, the cerebellum, and the thalamus.
Afterwards, \textcite{antogniniIsofluraneAnesthesiaBlunts1997} transformed the
data into blood oxygenation level dependent (BOLD) signals for each region of
interest by computing the proportion of active voxels within each region and
applying an arcsine transformation. For illustration purposes, we focus only on
the cortex and cerebellum and model their latent activation using two state
variables, which allows us to visualize the complete learned dynamics.
Additionally, to keep the example simple, we do not explicitly model the
external stimulus.

\begin{figure}[!ht]
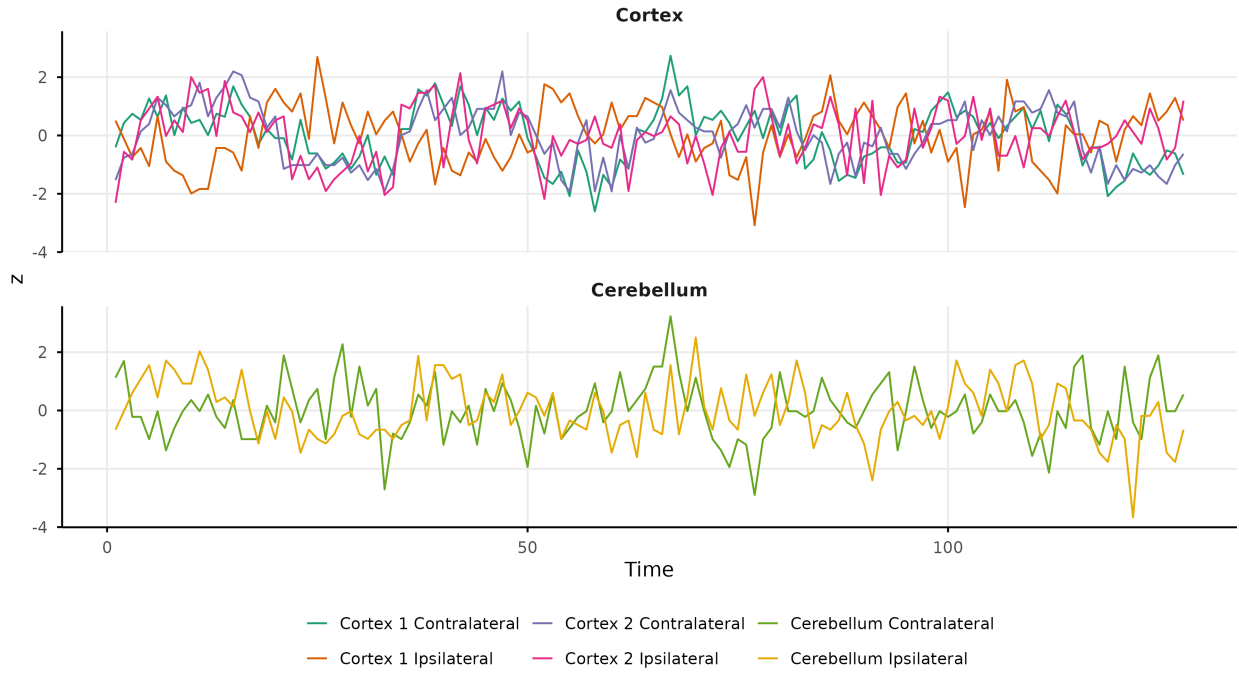

  \caption{Standardized fMRI BOLD signals}
  \fitfigure{fmri_data.png}
  \figurenote{This figure shows the standardized BOLD signals from the
  bilateral primary and secondary somatosensory cortices and cerebellum.}
  \label{fig:fmri_data}
\end{figure}

Figure~\ref{fig:fmri_data} shows the standardized BOLD signals from the
bilateral primary and secondary somatosensory cortices and cerebellum. The BOLD
signal is an indirect indicator of neural activity, as increases in local blood
oxygenation are associated with neural activation. Particularly, the BOLD
signals from the somatosensory cortices show a clear response to the tactile
stimulus. In three regions, the BOLD signal increases when the stimulus is
applied and decreases when it is removed, whereas the remaining region shows the
opposite pattern. This change in the BOLD signals reflects the hemodynamic
response to the stimulus \parencite{jahnAndrewjahnAndysBrainBook2022}, which can
be modelled using SSMs
\parencite{rieraStatespaceModelHemodynamic2004,hongStatespaceModelsImpulse2014}.

The goal of this example is not to provide a detailed model of the
neurophysiology underlying the hemodynamic response
\parencite{fristonNonlinearResponsesFMRI2000}, but rather to illustrate how the
GP-SSM can capture the switching activation patterns between periods of
stimulation and rest. The CFA measurement model is particularly useful for this
data set, because some participants exhibit the same activation patterns across
contralateral and ipsilateral regions, whereas others show reversed activation
patterns in some of the ipsilateral regions. Within the CFA model, this can be
accounted for by positive and negative factor loadings, respectively.

To this end, we use a model with two latent states and constrain the loading
matrix in such a way that all cortical regions of interest load on one state
variable and all cerebellar regions load on the other. The first indicator for
each state is used as an anchor variable. The resulting state variables could,
for example, be interpreted as latent activation factors underlying the observed
cortical and cerebellar BOLD signals. The remainder of the model was specified
analogously to the previous example, except that priors were added for the free
loading parameters and the prior for the measurement error covariance was
adjusted to account for the larger number of indicators

\begin{equation}\label{eq:fmri_priors}
  \begin{aligned}
    \bm{\lambda}  & \sim \N{\bm{0}}{2 \, \identity[4]}, \qquad
    \Sigmaepsilon & \sim \IW{11}{\identity[6]}.
  \end{aligned}
\end{equation}

\noindent The model was fitted using the same procedure as in the previous
example and converged at every iteration of the tuning algorithm. The final
model used 14 basis functions for the cortical activation dimension and 12 basis
functions for the cerebellar activation dimension.

Prior and posterior predictive checks are reported in Appendix~\ref{app:B}.
Figure~\ref{fig:fmri_prediction} shows the posterior predictive distribution for
all observed brain regions as a time series together with the smoothing
distributions of the two latent states each region is loading on. Overall, the
posterior predictive distribution reproduces the main patterns in the data
sufficiently well, which indicates that the GP-SSM fits the observed BOLD
signals adequately. The smoothing distribution of the cortical activation state
shows clear switching between periods of high and low activation, whereas the
cerebellar activation state shows a more gradual oscillatory behavior.

Note, that the sign of each latent state is determined by the contralateral
region that functions as its anchor variable.
For the somatosensory cortices, this is unproblematic because cortical
representations are typically crossed, meaning that the strongest response to
the tactile stimulus would be expected in the contralateral cortex. In the
cerebellum, however, the strongest response is expected in the ipsilateral
region. Indeed, Figure~\ref{fig:fmri_prediction} shows that
the ipsilateral cerebellar signal is in phase with the administered stimulus,
whereas the contralateral signal is reversed. Because the
cerebellar state is anchored to the contralateral cerebellum, its sign is
reversed relative to the stimulus response. As a consequence, low values 
of the cerebellar state may be interpreted as a high cerebellar stimulus 
response, and vice versa. 

\begin{figure}[!ht]
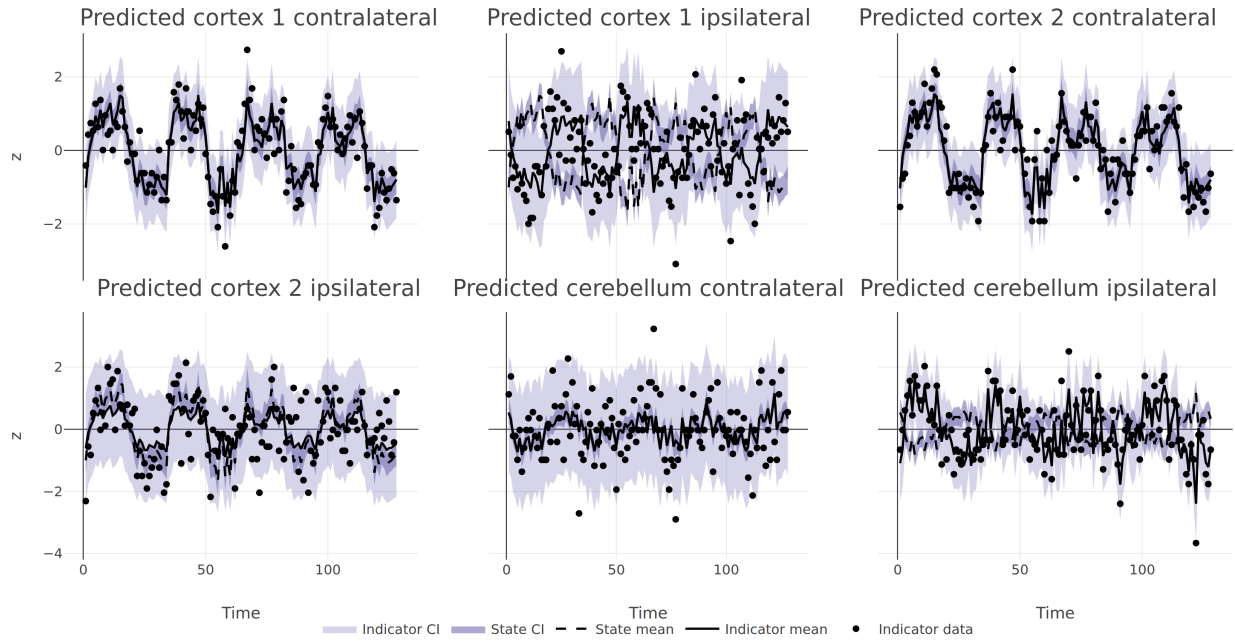

  \caption{Posterior predictive and smoothing distribution}
  \fitfigure{fmri_prediction.png}
  \figurenote{This figure shows the posterior predictive distributions for the
  measured brain regions together with the smoothing distributions of the
  cortical and cerebellar activation states. The black line corresponds to the
  posterior predictive mean and the lighter shaded area to its
  95\% credible interval. The dotted line indicates the posterior mean of the
  smoothing distribution and the darker shaded area its 95\% credible
  interval.}
  \label{fig:fmri_prediction}
\end{figure}

Table~\ref{tab:fmri_post_pars} summarizes the posterior distributions of all
model parameters. The posterior mean of the lengthscale parameter is $1.10$,
which indicates that the learned dynamic in this example is varying faster
across the state-space than in the predator-prey example. Additionally, the
loadings for the ipsilateral primary somatosensory cortex and ipsilateral
cerebellum are negative, confirming that their activation patterns are indeed
reversed relative to the latent states they load on.

\begin{table}[!ht]
  \caption{Posterior parameter summaries from the fMRI model}
\centering
\begin{tabular}{rlrrrrrrr}
  \hline
 & Variable & Mean & SD & Q5 & Q95 & R-hat & ESS Bulk & ESS TAIL \\ 
  \hline
1 & alpha & 1.23 & 0.37 & 0.79 & 1.90 & 1.00 & 7543.56 & 7987.78 \\ 
  2 & rho & 1.10 & 0.26 & 0.74 & 1.54 & 1.00 & 6374.67 & 7016.72 \\ 
  3 & Sigma\_eta[1,1] & 0.21 & 0.04 & 0.15 & 0.28 & 1.00 & 4406.27 & 5873.44 \\ 
  4 & Sigma\_eta[1,2] & -0.02 & 0.02 & -0.05 & 0.01 & 1.00 & 5619.75 & 6817.29 \\ 
  5 & Sigma\_eta[2,2] & 0.10 & 0.03 & 0.06 & 0.16 & 1.00 & 1257.51 & 2759.01 \\ 
  6 & Lambda[2,1] & -0.80 & 0.11 & -0.99 & -0.62 & 1.00 & 3027.47 & 4275.93 \\ 
  7 & Lambda[3,1] & 1.03 & 0.09 & 0.89 & 1.18 & 1.00 & 2802.99 & 3931.51 \\ 
  8 & Lambda[4,1] & 0.59 & 0.12 & 0.40 & 0.78 & 1.00 & 3024.51 & 4905.86 \\ 
  9 & Lambda[6,2] & -2.10 & 0.42 & -2.82 & -1.45 & 1.00 & 942.89 & 1914.42 \\ 
  10 & Sigma\_epsilon[1,1] & 0.24 & 0.05 & 0.17 & 0.33 & 1.00 & 2934.19 & 5214.33 \\ 
  11 & Sigma\_epsilon[1,2] & 0.16 & 0.04 & 0.09 & 0.23 & 1.00 & 4992.06 & 6265.41 \\ 
  12 & Sigma\_epsilon[2,2] & 0.54 & 0.08 & 0.41 & 0.68 & 1.00 & 4552.07 & 5895.66 \\ 
  13 & Sigma\_epsilon[1,3] & -0.04 & 0.03 & -0.09 & 0.01 & 1.00 & 3765.52 & 5344.10 \\ 
  14 & Sigma\_epsilon[2,3] & 0.09 & 0.04 & 0.02 & 0.16 & 1.00 & 4149.32 & 6123.54 \\ 
  15 & Sigma\_epsilon[3,3] & 0.23 & 0.05 & 0.16 & 0.31 & 1.00 & 3664.53 & 4779.20 \\ 
  16 & Sigma\_epsilon[1,4] & -0.07 & 0.05 & -0.15 & 0.02 & 1.00 & 4152.27 & 5911.12 \\ 
  17 & Sigma\_epsilon[2,4] & -0.01 & 0.06 & -0.11 & 0.10 & 1.00 & 4930.18 & 6650.96 \\ 
  18 & Sigma\_epsilon[3,4] & -0.02 & 0.05 & -0.11 & 0.07 & 1.00 & 3287.25 & 5590.19 \\ 
  19 & Sigma\_epsilon[4,4] & 0.72 & 0.10 & 0.57 & 0.89 & 1.00 & 5888.47 & 7259.63 \\ 
  20 & Sigma\_epsilon[1,5] & 0.11 & 0.06 & 0.02 & 0.21 & 1.00 & 4582.94 & 6480.24 \\ 
  21 & Sigma\_epsilon[2,5] & 0.10 & 0.07 & -0.01 & 0.22 & 1.00 & 5855.34 & 7119.11 \\ 
  22 & Sigma\_epsilon[3,5] & 0.03 & 0.05 & -0.06 & 0.12 & 1.00 & 3741.15 & 5495.88 \\ 
  23 & Sigma\_epsilon[4,5] & 0.11 & 0.08 & -0.01 & 0.24 & 1.00 & 6033.37 & 7013.88 \\ 
  24 & Sigma\_epsilon[5,5] & 0.85 & 0.12 & 0.67 & 1.06 & 1.00 & 4869.02 & 5968.15 \\ 
  25 & Sigma\_epsilon[1,6] & 0.01 & 0.04 & -0.06 & 0.08 & 1.00 & 4004.60 & 5768.73 \\ 
  26 & Sigma\_epsilon[2,6] & -0.05 & 0.06 & -0.14 & 0.04 & 1.00 & 4716.11 & 6657.58 \\ 
  27 & Sigma\_epsilon[3,6] & 0.00 & 0.04 & -0.06 & 0.07 & 1.00 & 4865.33 & 6441.92 \\ 
  28 & Sigma\_epsilon[4,6] & -0.01 & 0.06 & -0.11 & 0.09 & 1.00 & 4365.70 & 6263.56 \\ 
  29 & Sigma\_epsilon[5,6] & 0.30 & 0.07 & 0.18 & 0.42 & 1.00 & 3050.02 & 5140.21 \\ 
  30 & Sigma\_epsilon[6,6] & 0.27 & 0.08 & 0.15 & 0.42 & 1.00 & 1689.40 & 3385.87 \\ 
  31 & log\_lik & -639.67 & 36.67 & -701.66 & -581.21 & 1.00 & 2741.92 & 4281.44 \\ 
   \hline
\end{tabular}

  \label{tab:fmri_post_pars}
\end{table}

\begin{figure}[!ht]
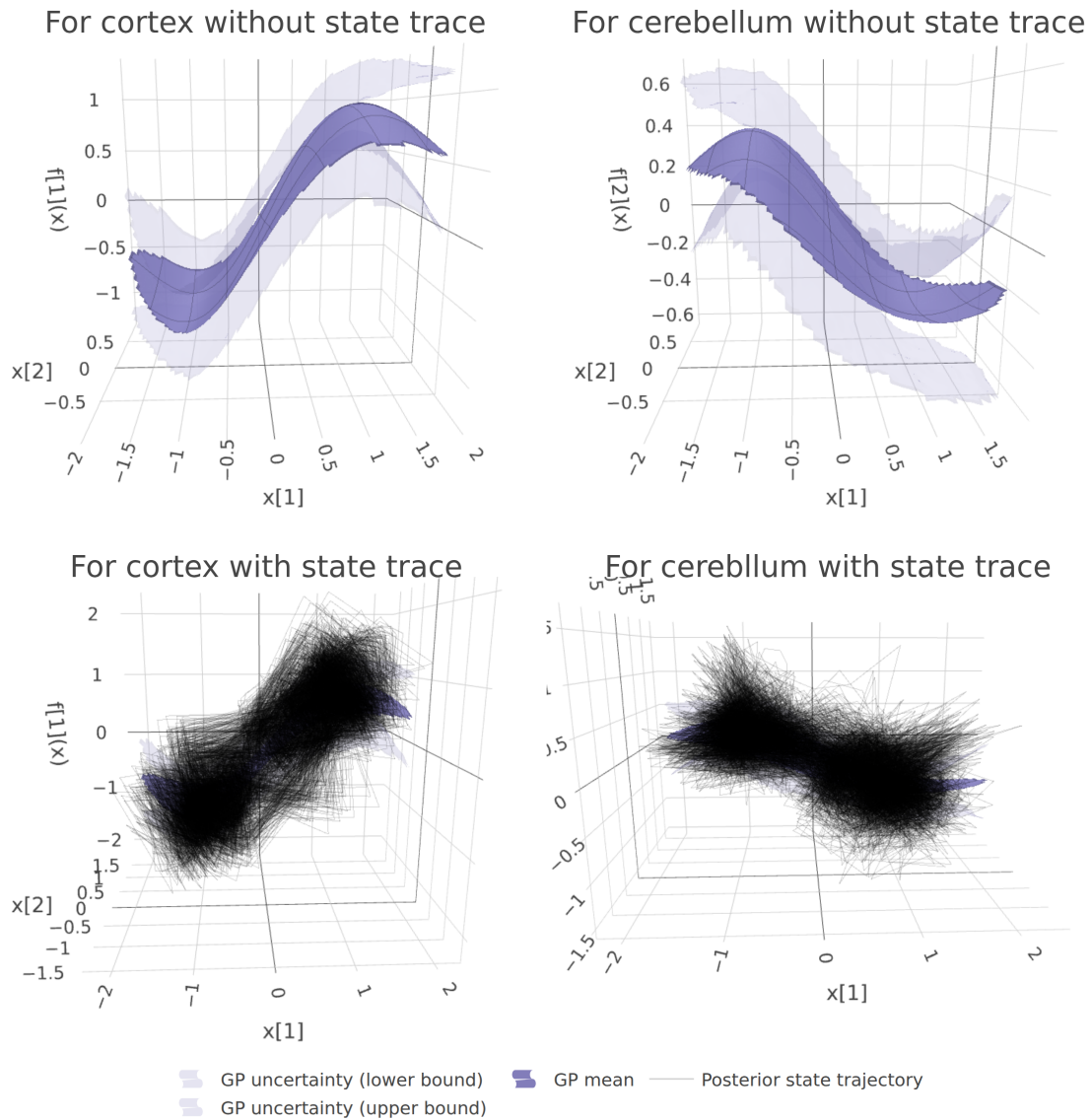

  \caption{Posterior Gaussian process surface}
  \fitfigure{fmri_posterior_gp.png}
  \figurenote{This figure shows the posterior GP for the fMRI activation model.
  Here, $x_1$ corresponds to the latent cortical activation state and $x_2$ to
  the latent cerebellar activation state. The left panels show the posterior GP
  for predicting cortical activation at the next time point and the right panels
  show the corresponding prediction surface for cerebellar activation. The dark
  surface corresponds to the posterior mean and the lighter surfaces the
  posterior 95\% credible interval. The lower panels additionally display 100
  sampled trajectories from the smoothing distribution. Note that the different
  posterior GP surfaces are not depicted using the same orientation or scale, as
  each surface is shown from the perspective that best illustrates its
  features.}
  \label{fig:fmri_posterior_gp}
\end{figure}

Figure~\ref{fig:fmri_posterior_gp} shows the posterior GP. The left panels show
the prediction surface for the cortical activation state (i.e., $x_1$), whereas
the right panels display the prediction surface for the cerebellar activation
state (i.e., $x_2$). Compared to the predator-prey example, the posterior GP
varies more rapidly over the supported region of the state space and appears
more nonlinear. This highlights the benefit of estimating the dynamic model
non-parametrically instead of imposing a linear dynamic model, which would be
likely misspecified in this case. The lower panels of
Figure~\ref{fig:fmri_posterior_gp} additionally show samples from the smoothing
distribution, which appear to cluster around two distinct points. 

\begin{figure}[!ht]
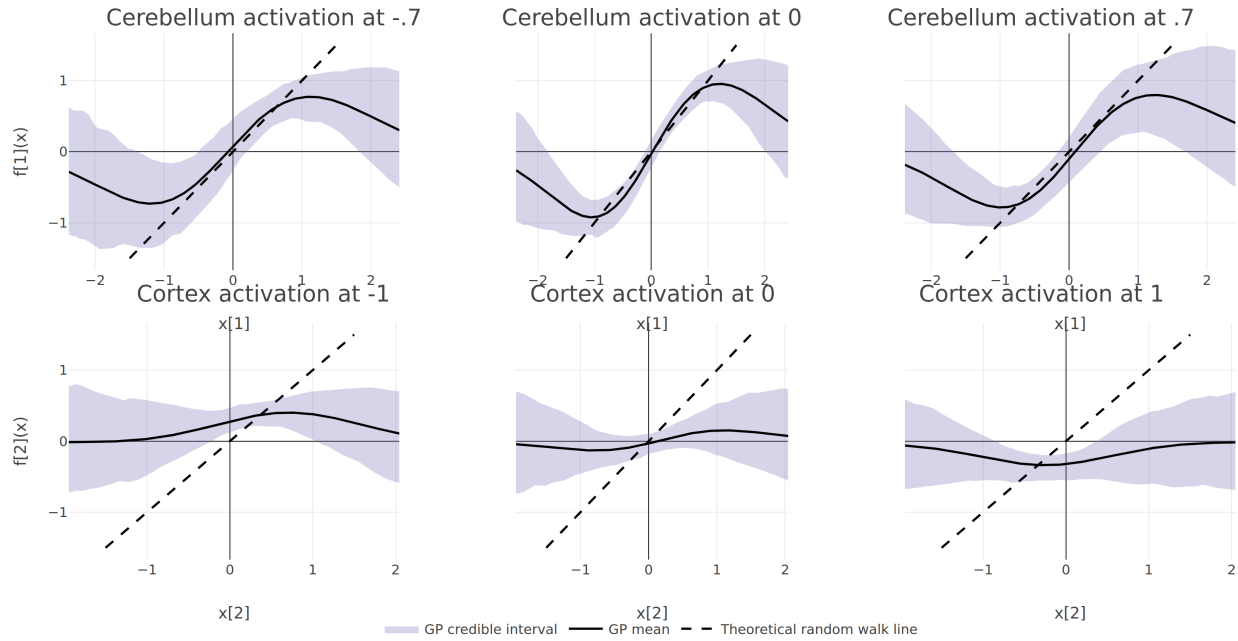

  \caption{Conditional slices through the posterior Gaussian process surfaces}
  \fitfigure{fmri_slices.png}
  \figurenote{This figure shows conditional slices through the posterior GP.
  Each panel displays the dynamics of one state while holding the other state
  fixed at a low, medium, or high value. The dashed line corresponds to
  $f(x)=x$, such that intersections between the posterior GP and this line
  correspond to conditional equilibrium points.}
  \label{fig:fmri_slices}
\end{figure}

To further interpret the learned dynamics, Figure~\ref{fig:fmri_slices} shows
conditional slices through the posterior GP. Each slice describes the implied
dynamics of one state while holding the other state fixed at a low, medium, or
high value. The dashed line corresponds to $f(x)=x$. When the posterior GP lies
on this line, the predicted state value is equal to the current state value.
This means that the conditional dynamics of the state would equal a random walk.
If the GP lies above this random walk line, the state is expected to increase,
whereas if the GP lies below this line it is expected to decrease. As a result,
any point where the GP crosses this line from above becomes a stable equilibrium
and any point where it crosses from below becomes an unstable equilibrium.   

The conditional dynamics of the cortical activation state show a single 
high activation equilibrium point when the cerebellar activation is also 
high (i.e., when the value of the cerebellar state is low). 
As the cerebellar activity decreases, the conditional dynamics of the
cortical activation state result in two stable conditional equilibria, 
one with high activation and one with low activation, and an unstable 
conditional equilibrium in between. For low levels of cerebellar activity, 
the conditional dynamics of the cortical activation show a single 
low activation equilibrium point. The conditional dynamics of the 
cerebellar activation state have a single stable equilibrium whose 
location depends on the level of cortical activation. 

\begin{figure}[!ht]
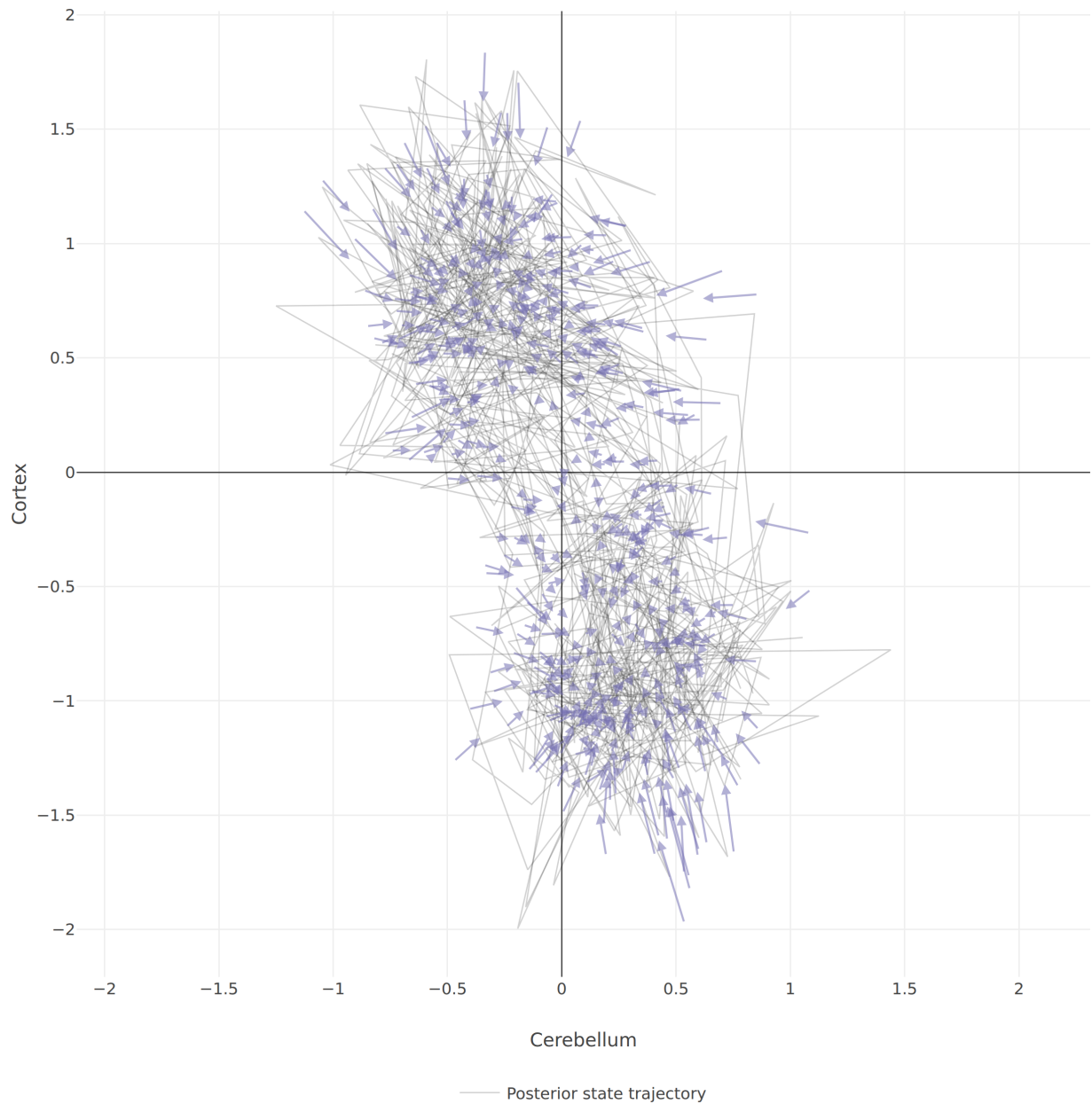

  \caption{Empirical vector field}
  \fitfigure{fmri_vector_field.png}
  \figurenote{This figure shows the empirical vector field learned by the
  GP-SSM. Each arrow starts at a sample from the smoothing distribution and
  points toward the state predicted by the posterior GP at the next time point
  at 25\% scale. Overlaid are five traces from the smoothing distribution. This
  figure illustrates how the learned dynamic system evolves from different
  locations in the state space.}
  \label{fig:fmri_vector_field}
\end{figure}

Taken together, the learned dynamics are consistent with a bistable system that
is primarily driven by the bistability of the cortical activation state. This is
particularly visible in the learned empirical vector field depicted in
Figure~\ref{fig:fmri_vector_field}, which shows that the state trajectories tend
to move towards two stable regions. When interpreting the observed bistability 
the external stimulus should be taken into account. Because the stimulus that
is driving the switching between high and low levels of cortical activation is
not included in the model, the GP-SSM must explain both periods of high and
low activation through its internal dynamics, which leads to the appearance of
two stable activation regimes\footnote{ If the stimulus is explicitly modelled
as a time-varying covariate in the GP-SSM, the high activation equilibrium is
largely explained by the stimulus and the learned dynamics show only the low
activation equilibrium.}. 

\section{Discussion}

In the present paper, we described the GP-SSM
\parencite{frigolaBayesianInferenceLearning2013,turnerStateSpaceInferenceLearning2010}
and showed how it can be combined with an identified and structured CFA
measurement model \parencite{muthenBayesianStructuralEquation2012}. We further
demonstrated how the resulting model can be estimated efficiently using the HSGP
approximation \parencite{solinHilbertSpaceMethods2019,
riutort-mayolPracticalHilbertSpace2023, mukherjeeHilbertSpaceMethods2025}. To
this end, we proposed two modifications to the Gibbs sampler from
\textcite{svenssonComputationallyEfficientBayesian2016}, which substantially
improved its sampling efficiency and convergence. This can be seen in
Figure~\ref{fig:orig_trace}, where we provide MCMC chains from our sampler for
the same dataset used by \textcite{svenssonComputationallyEfficientBayesian2016}
as a direct comparison.

The proposed sampler converged reliably for nearly all simulated datasets in the
SBC. Furthermore, the SBC showed no evidence that the proposed version of the
GP-SSM or its inference procedure were miscalibrated
\parencite{modrakSimulationBasedCalibrationChecking2022}. Moreover, the results
indicate that fitting a HSGP-SSM produces well-calibrated posterior draws for
the corresponding GP-SSM. To our knowledge, this is the first Bayesian estimator
for the GP-SSM that has been systematically validated on many datasets and can
be readily applied in substantive research. This is an important step towards
making GP-SSMs more accessible in practice. Nonlinear SSMs are notoriously
challenging to estimate using conventional MCMC methods \parencite[e.g., the
Hamiltonian Monte Carlo sampler implemented in
STAN][]{hoffmanNoUTurnSamplerAdaptively2014}, because the nonlinear dynamics can
induce strong dependencies and multimodality in the smoothing distribution of
the state variables. This makes SMC based methods, such as the one presented in
this paper, one of the most promising approaches for the exact Bayesian
inference of nonlinear state-space models
\parencite{sarkkaBayesianFilteringSmoothing2023,chopinIntroductionSequentialMonte2020}.

Beyond the simulation study, we illustrated how the GP-SSM can be used to learn
nonlinear dynamic systems directly from empirical data. This allows researchers
to capture complex nonlinear phenomena in a data-driven manner rather than
relying on potentially misspecified parametric models. The empirical examples
also highlight the utility of using an identified measurement
model, which yields directly interpretable posterior inferences for the
model parameters, the GP, and state variables, which can in turn be used to 
inform substantive theory and guide the development of future parametric
models.

\subsection{The curse of dimensionality}

The main limitation of the proposed approach is that it is unlikely to scale
well to larger numbers of state variables, as it is affected by the curse of
dimensionality in several ways \parencite{altmanCursesDimensionality2018}.

First, the number of particles that are required for the PGAS to be efficient
tends to increase with the number of state variables. As the number of state
variables grows, more particles are needed to approximate the smoothing
distribution with the same precision. Although, in theory, the PGAS can target
the correct smoothing distribution using at least two particles, using too few
particles can substantially reduce its mixing properties and slow down the
convergence of the sampler \parencite{andrieuParticleMarkovChain2010,
chopinParticleGibbsSampling2015}. In our testing, using too few particles caused
convergence problems for the state variables and lead to a downward bias in the
estimated dynamic error variance when the smoothing distribution was not
sufficiently explored. We therefore recommend to carefully monitor the
convergence of the sampler and increase the number of particles should the
effective sample size be too low.  

Second, the computational cost of the HSGP increases rapidly as the number of
state variables increases. The basis-function representation of a multivariate
kernel is given by the Cartesian product of the basis functions in each input
dimension, which causes the number of basis functions to grow exponentially with
the number of state variables
\parencite{riutort-mayolPracticalHilbertSpace2023}. This can quickly lead to
impractical computation times. In our simulation study, some models with two
state variables and a small lengthscale required nearly 2{,}000 basis
functions, resulting in computing times of up to four days.

One possible way to reduce this problem is through additive GPs. A very useful
property of GPs, from a modelling perspective, is that they are additive and
multiplicative, which allows researchers to construct more flexible and
fine-tuned models. This means that adding or multiplying GPs is equal to adding
or multiplying their respective mean and covariance functions
\parencite{rasmussenGaussianProcessesMachine2006}. The multivariate kernel used
in this paper can be seen as being multiplicative across all input dimensions,
allowing it to capture arbitrary nonlinear dependencies between the state
variables. While this provides a lot of flexibility, it comes at a substantial
computational cost in higher dimensions. A purely additive GP, in contrast,
assumes that the nonlinear effect of each input variable can be represented
separately, and therefore it cannot capture any nonlinear interactions between
the state variables \parencite{duvenaudAdditiveGaussianProcesses2011a}. However,
when represented through basis functions additive models require considerably
less basis functions and are less expensive to estimate
\parencite{woodGeneralizedAdditiveModels2006}. An ideal model would therefore
only use multiplicative kernels where they are necessary to capture interactions
in the data while preserving an additive structure elsewhere
\parencite{duvenaudAdditiveGaussianProcesses2011a,
duvenaudStructureDiscoveryNonparametric}. In addition to their computational
advantages, additive models are also often easier to interpret.

A related limitation is that our formulation of the GP-SSM uses the same GP
prior for the dynamics of all state variables, which is necessary for the
conjugacy of the Gibbs sampler
\parencite{svenssonComputationallyEfficientBayesian2016}. While the posterior
dynamics of the states can still differ, this assumption can become problematic
when the different state dynamics require different kernels, hyperparameters, or
input structures. \textcite{mishraDomainAwareGaussianProcess2026} provide an
alternative to this, which makes it possible to use different GP priors for each
state dynamic by assuming independent dynamic errors. This splits the
multivariate regression problem into separate univariate regressions that can be
estimated separately. However, this can also become problematic, when the
dynamic errors are indeed correlated. This can, for example, happen when
multiple state variables are affected by the same external influences, as
illustrated in the predator-prey example. To our knowledge, there is currently
no Bayesian estimator for the HSGP-SSM that allows for state specific GP priors
and a fully unrestricted dynamic error covariance matrix. 

Lastly, interpreting the learned dynamics becomes increasingly difficult in
higher dimensions. In the empirical examples, we have focused on systems with
two state variables, which made it possible to visualize the complete dynamics
directly. This is generally not possible for systems with more than two state
variables. Nevertheless, there are several graphical tools available to analyze
and interpret these higher dimensional systems. As illustrated, it is possible
to visualize conditional slices through the learned dynamics of one state
variable, while holding the other state variables constant at specific values.
This can provide insights into conditional behavior of individual state
variables in specific regions of the state space. Another option is to analyze
the learned dynamics in the phase-space
\parencite{cuiAnalyzingMultidimensionalFormal2026}, through the learned
empirical vector fields. For systems with up to three state variables, these
vector fields can be visualized directly and for higher-dimensional systems,
MCMC sampling makes it easy to marginalize over selected state variables and
project the phase space onto a lower-dimensional plane. For systems with many
state variables, it might also be useful to fix all state variables to specific
values and visualize the partial derivatives of the posterior GP as a
linearization network
\parencite{cuiExaminingFeasibilityNonlinear2024,krocCaseCurveParametric2025}.
This makes it possible to study how a system with many state variables is
expected to change from a particular point in the state space.  While these
tools can provide valuable insights into the learned dynamic system, reasoning
from conditional dynamics and projections back to the behavior of the full
dynamic system becomes increasingly difficult as its dimensionality increases.

\subsection{Future directions}

One of the main strengths of the GP-SSM is its flexibility
\parencite{frigola-alcaldeBayesianTimeSeries2015}. In the present paper, we
assumed a zero mean function and a squared-exponential kernel for simplicity.
However, the proposed method can readily accommodate other mean functions that
are either fixed or regression models with Gaussian priors, which can directly
be incorporated into the HSGP basis-function expansion. Similarly, other
stationary covariance kernels, or combinations thereof, can be used in place of
the squared-exponential kernel \parencite{solinHilbertSpaceMethods2019,
riutort-mayolPracticalHilbertSpace2023}. Different kernels can be used to encode
different prior assumptions about the form of the system dynamics, such as their
periodicity and roughness \parencite{rasmussenGaussianProcessesMachine2006,
duvenaudAutomaticModelConstruction}. It is also straightforward to use separate
lengthscales for each input dimension. Such alternatives may be particularly
useful because a zero mean GP with a squared exponential kernel tends to revert
towards zero in regions that are only weakly informed by the data. As a result,
the learned dynamics may curve towards zero near the boundaries of the observed
data, as is the case in the fMRI example, which can make it more difficult to
interpret them.

The Gibbs sampler presented in this paper can also accommodate other basis
function constructions. Since the HSGP representation corresponds to a
multivariate basis function regression with matrix normal regression weights,
the basis functions can, in principle, be replaced with other basis functions
with Gaussian priors \parencite{svenssonFlexibleStateSpace2017}. The GP
construction then mainly provides a principled way to derive basis functions
together with prior distributions that impose an appropriate degree of
regularization. One possible alternative could be Bayesian smoothing splines
\parencite{millerBayesianViewsGeneralized2021,
sorensenModelingCyclesTrends2025}, although this would additionally require the
selection of suitable knot locations in the latent state space.

In addition to this, it would be useful to extend the presented approach to
different measurement models. The CFA measurement model assumes linear Gaussian
measurements and a known measurement structure, which may not be appropriate in
some practical applications. Extending the GP-SSM to alternative measurement
models is conceptually straightforward
\parencite{frigolaBayesianInferenceLearning2013}, but extending the Gibbs
sampler is more restrictive. In addition to the usual conjugacy requirements of
the Gibbs sampler, the optimal proposal distribution for the particle filter
requires access to the local posterior distribution of the state variables
conditional on the current observations
\parencite{snyderParticleFiltersOptimal2011}. When this local conjugacy is not
directly satisfied by the measurement model, one possible solution is to
introduce additional Gaussian latent variables
\parencite{muthenBayesianStructuralEquation2012}, which separate the particle
filter from the non-Gaussian measurement model. A particularly interesting 
measurement model to use as part of the GP-SSM could be the 
exploratory factor analysis model, which learns the structure of the loading 
matrix in a data driven manner rather than assuming that it is known 
\parencite{contiBayesianExploratoryFactor2014}.

Another direction could be to extend the presented GP-SSM to hierarchical or
multilevel data structures. In empirical applications, researchers frequently
collect data from multiple related dynamic systems. In psychology, for example,
data are often collected from multiple individuals who are assumed to exhibit
similar, but not identical, dynamics
\parencite{jongerlingMultilevelAR1Model2015,mulderThreeExtensionsRandom2021}.
Existing hierarchical and mixture state-space models can capture the
heterogeneity between these systems while simultaneously capitalizing on their
similarities by pooling information across them
\parencite{liuMixedEffectsStateSpace2011, hunterStateSpaceMixture2024,
driverHierarchicalBayesianContinuous2018, reinThreeStepStateSpace2026}, which
can improve the accuracy of parameter estimates and reduce the number of
observations required for reliable inference from each individual system
\parencite{veenmanBayesianHierarchicalModeling2023}. This might be particularly
beneficial for GP-SSMs, since learning the nonlinear dynamics likely requires
more observations than estimating a fixed parametric model. Hierarchical
extensions of GPs already exist \parencite{karchGaussianProcessPanel2020,
hoffmannComputationallyEfficientMultilevel2026,
timonenEmphLgprInterpretableNonParametric2021}. However, it is currently unclear
how these hierarchical GP models can be represented within the HSGP
approximation and integrated in the Gibbs sampling scheme proposed in this
paper.

Lastly, there remains the issue of specifying good prior distributions for the
hyperparameters. Priors that are too diffuse may not impose a sufficient amount
of regularization on the model, which can cause identifiability and convergence
problems during the MCMC sampling
\parencite{betancourtRobustGaussianProcess2020}. Conversely, overly informative
priors can restrict the flexibility of the GP and may prevent it from accurately
fitting the data. This problem is only exacerbated, when the GP uses latent
variables as inputs. This can create strong dependencies between the prior
distributions of the hyperparameters and the error covariance matrices, such
that using incompatible priors can lead to convergence issues and undesirable
model behaviors \parencite{mukherjeeHilbertSpaceMethods2025}. We therefore
recommend carefully specifying the hyperparameter priors when fitting GP-SSMs,
using prior predictive checks to assess their implications and monitoring their
effect on the model convergence and fit
\parencite{gabryVisualizationBayesianWorkflow2019}. More generally, further
research is needed to study the sensitivity of the GP-SSM to different prior
distributions and develop robust default recommendations. 

\noindent

\printbibliography[]

\appendix

\section{}\label{app:A}

\begin{figure}[!ht]
  \caption{ECDF differences for the univariate models in the additional simulation 1}
  \fitfigure{ecdf_diff_1.png}
  \figurenote{This figure shows the difference between the ECDF of the ranks
    for each parameter and a theoretical uniform CDF. The blue area represents a
    $95\%$ confidence region for the largest deviation from the uniform CDF. If
    the sampler and model are well calibrated, the difference between the sample
    ECDF and the uniform CDF should remain within this region for $95\%$ of the
    parameters.}
  \label{fig:ecdf_diff_1}
\end{figure}

\begin{figure}[!ht]
  \caption{ECDF differences for the univariate models in the additional simulation 2}
  \fitfigure{ecdf_diff_2.png}
  \figurenote{This figure shows the difference between the ECDF of the ranks
    for each parameter and a theoretical uniform CDF. The blue area represents a
    $95\%$ confidence region for the largest deviation from the uniform CDF. If
    the sampler and model are well calibrated, the difference between the sample
    ECDF and the uniform CDF should remain within this region for $95\%$ of the
    parameters.}
  \label{fig:ecdf_diff_2}
\end{figure}

\begin{figure}[!ht]
  \caption{ECDF differences for the univariate models in the additional simulation 3}
  \fitfigure{ecdf_diff_3.png}
  \figurenote{This figure shows the difference between the ECDF of the ranks
    for each parameter and a theoretical uniform CDF. The blue area represents a
    $95\%$ confidence region for the largest deviation from the uniform CDF. If
    the sampler and model are well calibrated, the difference between the sample
    ECDF and the uniform CDF should remain within this region for $95\%$ of the
    parameters.}
  \label{fig:ecdf_diff_3}
\end{figure}

\section{}\label{app:B}

\begin{figure}[!ht]
  \caption{Prior and posterior predictive checks for the predator-prey model}
  \fitfigure{lv_ppcheck.png}
  \label{fig:pp1}
\end{figure}

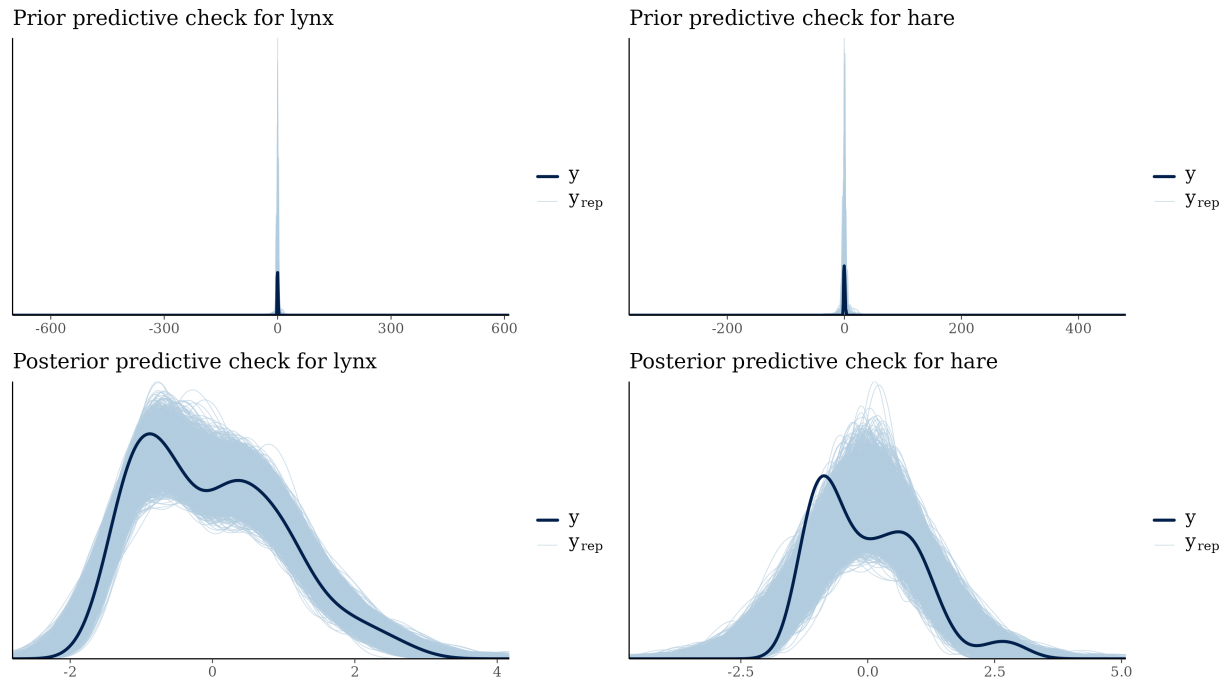
\begin{figure}[!ht]
  \caption{Prior predictive checks for the fMRI model}
  \fitfigure{prior_pcheck.png}
  \label{fig:pp2}
\end{figure}

\begin{figure}[!ht]
  \caption{Posterior predictive checks for the fMRI model}
  \fitfigure{posterior_pcheck.png}
  \label{fig:pp3}
\end{figure}

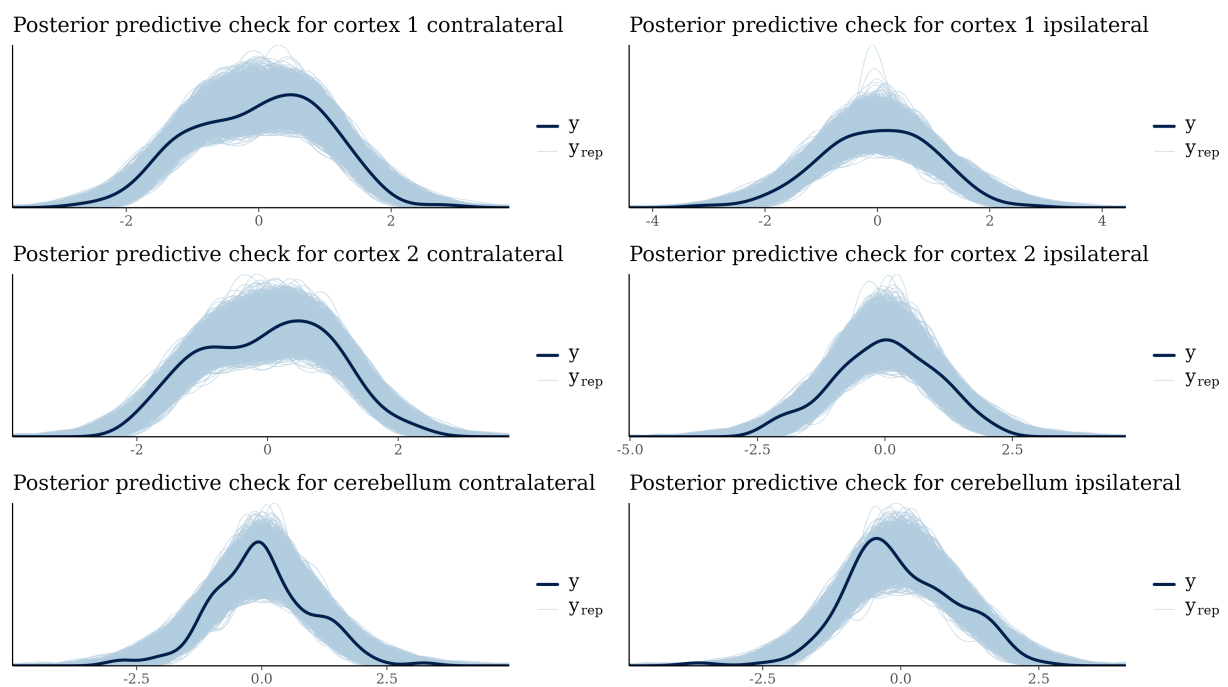
\end{document}